\documentclass[twocolumn,english,amsart,showpacs,preprintnumbers,amsmath,amssymb,floatfix,nofootinbib]{revtex4-1}

\usepackage{enumitem}
\usepackage[T1]{fontenc}
\usepackage{adjustbox}
\usepackage[latin9,utf8]{inputenc}
\usepackage{xcolor}
\usepackage{array}
\usepackage{amstext}
\usepackage{graphicx}
\usepackage{epstopdf}
\usepackage{esint}
\usepackage{float}
 \usepackage{tabularx}
\usepackage{dcolumn}
\usepackage{bm}
\usepackage{subfigure}
\usepackage{color}
\usepackage{float}
\usepackage{multirow}
\usepackage{xspace}
\usepackage{adjustbox}

\usepackage{booktabs}
\usepackage{amsmath}
\usepackage{hhline,multirow,tabularx}  
\DeclareGraphicsExtensions{.pdf}
\usepackage{latexsym,amssymb}
\usepackage[mathscr]{eucal}
\usepackage{rotating}
\usepackage{appendix}
\usepackage{ulem}
\usepackage{tikz,xcolor}
\usepackage[colorlinks = true,
linkcolor = blue,
urlcolor  = blue,
citecolor = blue,
anchorcolor = blue,
linktocpage]{hyperref}
\definecolor{lime}{HTML}{A6CE39}
\DeclareRobustCommand{\orcidicon}{%
	\begin{tikzpicture}
		\draw[lime, fill=lime] (0,0)
		circle [radius=0.16]
		node[white] {{\fontfamily{qag}\selectfont \tiny ID}};
		\draw[white, fill=white] (-0.0625,0.095)
		circle [radius=0.007];
	\end{tikzpicture}
	\hspace{-2mm}
}
\foreach \x in {A, ..., Z}{%
	\expandafter\xdef\csname orcid\x\endcsname{\noexpand\href{https://orcid.org/\csname orcidauthor\x\endcsname}{\noexpand\orcidicon}}
}

\newcommand{\x}{\ensuremath{x}\xspace}

\makeatletter
\begin{document}

\title{QCD analysis of valence structure functions using deep inelastic lepton-nucleon scattering}

\author{Javad Shahrzad 
}  
\email[]{javadshahrzad@semnan.ac.ir}
\author{Ali Khorramian\orcidB{}}
  \email[]{Khorramiana@semnan.ac.ir}
\affiliation{Faculty of Physics, Semnan University, P. O. Box 35131-19111, Semnan, Iran}

\date{\today}

\begin{abstract}
A new ``$\mathtt{SK24}$'' non-singlet QCD analysis of the structure functions at the NNLO approximation is performed, utilizing the global fit of the data from various charged lepton scattering experiments. We extract the valence parton distribution functions (PDFs) and provide a parametrization of them, along with the correlated errors for a wide range of $x$ and $Q^2$. We compare valence PDFs and their uncertainties with those from different PDF sets provided by various groups. We also obtain valence PDFs and the strong coupling constant $\alpha_{s}(M_Z^2)$, taking into account the nuclear correction concerning large $x$ as well as the target mass correction (TMC) and higher twist (HT) effects at the NNLO. In the large $x$ region, we extract the higher twist contributions of $xF_3(x,Q^2)$, $F_2^p (x,Q^2)$, and $F_2^d(x,Q^2)$. We determine $\alpha_{s}(M_Z^2)$ without and with considering the TMC and HT corrections and perform a comparison with the world average of $\alpha_{s}(M_Z^2)$ and other reported results. The extracted results concerning valence PDFs with their uncertainties and $\alpha_{s}(M_Z^2)$ value agree with available theoretical models.

\end{abstract}
\maketitle

\section{Introduction}
Deep-inelastic scattering (DIS) of high-energy leptons off hadrons is a fundamental process for studying the
structure of hadrons in terms of their parton distribution functions (PDFs) \cite{Taylor:1991ew,Kendall:1991np,Friedman:1991nq}. 
For both nucleons and nuclei, the data obtained from electron and neutrino DIS play a significant role in modern determinations of PDFs, illuminating and decoding the internal structure of hadrons. These kinds of data have offered detailed insights into parton densities, revealing valuable information at both small and large values of the parton momentum fraction $x$ of the nucleon. 
The investigation will once again take the spotlight at the future Electron-Ion Collider (EIC) \cite{Accardi:2012qut,AbdulKhalek:2021gbh}, proposing an opportunity to study DIS off nucleons and various nuclear targets with remarkable precision across a wide kinematic range, with particular emphasis on the large $x$ region.

Progress in understanding the inner structure of hadrons continues through ongoing research efforts, with new experiments, improved theoretical models, and increasingly sophisticated computational methods. Achieving precise theoretical calculations remains challenging despite having theoretically motivated expectations for certain aspects of the structure. Quantum chromodynamics (QCD) global fits to these data are instrumental in constraining the PDFs. DIS has been highly successful in investigating various features of QCD, including the study of polarized \cite{Khorramian:2020gkr, Nematollahi:2023dvj, Mirjalili:2022cal, Han:2021dkc, Salimi-Amiri:2018had, Sato:2016tuz, TaheriMonfared:2014var, Jimenez-Delgado:2013boa, Arbabifar:2013tma, Khorramian:2010qa, Leader:2010rb, Blumlein:2010rn, AtashbarTehrani:2007odq, Khorramian:2004ih} and unpolarized PDFs \cite{Yan:2022pzl, Gao:2022uhg, Hou:2022onq, Abdolmaleki:2019tbb, NNPDF:2021njg, Bailey:2020ooq, H1:2015ubc, Hou:2019qau} within a hadron. 

Furthermore, there has been a growing interest in the large $x$ behavior of proton PDFs in recent years \cite{Courtoy:2020fex, Ball:2016spl, Fu:2023rrs}. 
The importance of large $x$ PDFs is notable in specific high-energy scattering processes, such as the investigation of intrinsic heavy quarks within the proton, which is dominant at large $x$ \cite{BHPS1, BHPS2, Brodsky:2015fna, Brodsky:2020hgs}. In recent years, there have been increasing efforts to examine the role of intrinsic heavy quarks within a proton \cite{Ball:2022qks,Ball:2023anf}. The results presented in Ref.~\cite{Azizi:2022nqm} discuss the implications of the large $x$ PDFs on various high-energy processes.

Determining the valence PDFs with a particular emphasis on the large $x$ region is a topic of significant importance, where the impact of large $x$ PDFs is remarkable. Valence quarks, being among the constituent quarks, play a crucial role in understanding the internal structure of the proton. Very recently, new constraints on the $u$-valence PDF are reported in Ref. \cite{Aggarwal:2022cki}.

 Deep inelastic neutrino-nucleon scattering provides essential information for extracting  $xu_v(x,Q^2)$ and $xd_v(x,Q^2)$ valence quark densities. The non-singlet structure function, $xF_3(x,Q^2)$, plays an important role in the QCD global analysis of PDFs, especially at large $x$, where valence quark distributions are essential. The $xF_3$ structure functions of deep inelastic neutrino-nucleon scattering have been measured by several experimental groups, including the Chicago-Columbia-Fermilab-Rochester collaboration (CCFR) \cite{ccfr:1977}, Neutrinos at the Tevatron (NuTeV) \cite{Tzanov:2005kr}, CERN Hybrid Oscillation Research ApparatUS (CHORUS) collaboration at CERN \cite{Onengut:2005kv}, and CERN-Dortmund-Heidelberg-Saclay-Warsaw collaboration (CDHSW) \cite{Berge:1989hr}. These experimental data are a precise foundation for determining the valence quark densities and the strong coupling constant. Previous non-singlet analyses of CCFR data have been performed using various approximation methods based on orthogonal polynomial expansions, such as Jacobi polynomials \cite{Kataev:1994rj, Kataev:1996vu, Kataev:1997nc, Kataev:1997vv, Alekhin:1998df, Alekhin:1999af, Kataev:1999bp, Kataev:2001kk, Kataev:2002wr, Sidorov:2013aza, AtashbarTehrani:2009zz}, Bernstein polynomials \cite{Santiago:2001mh, Khorramian:2006wg}, and Laguerre polynomials \cite{GhasempourNesheli:2015tva}. Recent non-singlet QCD results based on CCFR, NuTeV, CHORUS, and CDHSW neutrino-nucleon data to determine $xu_v(x,Q^2)$ and $xd_v(x,Q^2)$, taking into account nuclear and higher twist corrections at the next-to-leading order (NLO) and next-to-next-to-leading order (NNLO) are presented in Ref.~\cite{Tooran:2019cfz} using the xFitter framework. 
The analysis covered a wide range of $x$ and $Q^2$, offering valuable insights into the valence quark distributions in the nucleon. The QCD analysis of the non-singlet $xF_3$ structure function, utilizing the Laplace transform approach, is reported in Ref. \cite{MoosaviNejad:2016ebo} without considering any nuclear and higher twist corrections.

In addition, performing a non-singlet QCD analysis of $F_2(x,Q^2)$ structure function using the world data for charged lepton scattering offers an opportunity to extract the valence quark parton densities and the associated correlated errors on a broad range of $x$ and $Q^2$. By employing non-singlet analysis of $F_2$, we can also extract the flavor non-singlet $xu_v(x,Q^2)$ and $xd_v(x,Q^2)$ PDFs from the available $e(\mu)p$ and $e(\mu)d$ world data of $F_2^p(x,Q^2)$ and $F_2^d(x,Q^2)$ in the valence-quark region, as well as from the $F_2^{NS}(x,Q^2)=2[F_2^p(x,Q^2)-F_2^d(x,Q^2)]$ structure functions. Data sets from various experiments, including BCDMS, NMC, SLAC, H1, and ZEUS, are utilized for this purpose \cite{BCDMS:1989ggw, BCDMS:1989qop, BCDMS:1989gtb, Whitlow:1991uw, NewMuon:1996fwh, H1:2000muc, H1:2003xoe, ZEUS:1998agx, ZEUS:2001mhd}. Our previous non-singlet analyses of electron(muon)-proton and deuteron $F_2$ data based on Jacobi polynomials are reported in Refs.~\cite{Khorramian:2008yh, Khorramian:2009xz}. Previous results of non-singlet analyses \cite{Blumlein:2006be}, presented without using the Jacobi polynomials approach, extended up to N$^3$LO, and instead, utilized the Mellin space technique. In fact, this analysis is based on the approach of J.~Bl\"umlein, et al. in Ref.~\cite{Blumlein:2006be} and has developed it by utilizing various data and different parameterization in $x$-space by QCDNUM open-source framework \cite{Botje:2010ay}.

\begin{figure*}[h!]

\includegraphics[width=1.\textwidth]{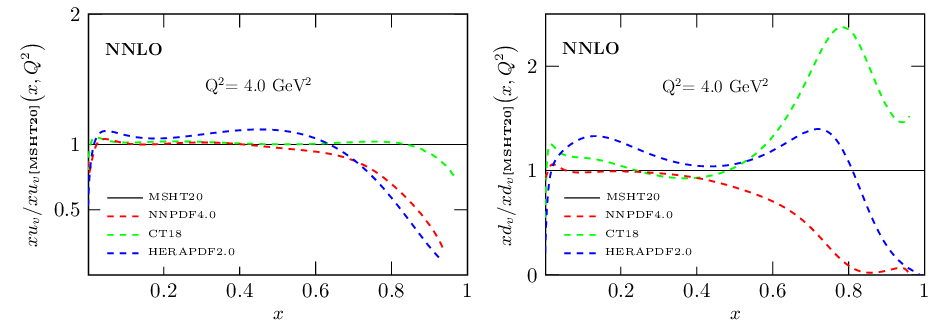}  
\caption{Valence PDF ratios of the NNLO  
NNPDF4.0 \cite{NNPDF:2021njg}, CT18 \cite{Hou:2019qau}, and HERAPDF \cite{H1:2015ubc} predictions to the MSHT20 set \cite{Bailey:2020ooq}, $xu_v/xu_{v}${\tiny[MSHT20]}$(x,Q^2)$,  as a function of $x$ at Q$^2= 4$ GeV$^2$.}
\label{fig:ModelRatio}
\end{figure*}

Our motivation to include $xF_3(x,Q^2)$ and $F_2(x,Q^2)$ data in our QCD non-singlet analysis is the study of valence PDFs at large $x$, where these kinds of PDF is not one of the very well known PDFs in the proton, particularly for $x$ > 0.5. In Figure~\ref{fig:ModelRatio}, we compare the NNLO valence PDF ratios of the NNPDF4.0 \cite{NNPDF:2021njg}, CT18 \cite{Hou:2019qau}, and HERAPDF2.0 \cite{H1:2015ubc} predictions to the MSHT20 set \cite{Bailey:2020ooq}. All PDF sets are imported from LHAPDF6 \cite{Buckley:2014ana}. This figure shows the sizable difference between the extracted $u$ and $d$ valence PDF results for large $x$ values. Therefore, in this study, we are motivated to provide a valence PDF set in the presence of $xF_3(x,Q^2)$ and $F_2(x,Q^2)$ data. We aim to find the influence of such data on the central value and corresponding uncertainties of valence PDFs, with a particular focus on large values of $x$ and the strong coupling constant. 

The current study explicitly emphasizes a non-singlet analysis to extract valence PDFs. It involves integrating DIS neutrino-nucleon $xF_3(x,Q^2)$ structure function data with $F_2(x,Q^2)$ structure function measurements obtained from charged electron (muon)-proton and deuteron collisions from various experiments. Indeed, this analysis benefits from being free of assumptions on the gluon PDF, resulting in improved systematic considerations. 
The reduced number of parameters needed to describe the parton distributions and the lack of correlation between the QCD strong coupling constant $\alpha_s$ and the gluon PDF contribute to the enhanced reliability of the results. Heavy flavor contributions for non-singlet structure function are investigated in Ref.~\cite{Blumlein:2006be}. As this article has pointed out, these effects are negligible at NNLO except for the large $Q^2$ region ($Q^2 \simeq 10^4$ GeV$^2$), which is not an effective area of the present paper. Additionally, according to Ref.~\cite{Blumlein:2021lmf}, it is demonstrated that within the large $Q^2$ region, heavy flavor corrections may reach up to 1\% for $x \leq 0.4$ at N$^3$LO level of accuracy. Moreover, it is observed that the relative effects tend to be larger at higher values of $x$.  In this analysis, $F_2$ also depends on sea quark corrections in the valence region to some extent. The $u$ and $d$-valence PDFs at $Q^2_0 = 4$ GeV$^2$ are affected only slightly, as demonstrated in Ref.~\cite{Blumlein:2012se}.

In this article, we manage our non-singlet QCD analysis using the QCDNUM open-source framework \cite{Botje:2010ay}. Incorporating nuclear corrections, target mass corrections (TMCs), and higher twist (HT) effects is essential to accommodate the diverse data sets utilized in this analysis. These corrections are performed during the fit procedure and play an essential role in our study. We incorporate these adjustments into our codes, which are not included in the main QCDNUM package.

The paper is organized as follows: We present a brief overview of the basic formalism for DIS structure functions and parton distributions in Section II. Moving on to Section III, we present the inputs for the fit, encompassing both theoretical and experimental aspects. This section also covers our parametrization for valence PDFs and the practical data sets utilized in the present QCD analysis. Section IV is dedicated to QCD corrections, where we discuss nuclear effects, target mass corrections, and higher twist effects, ensuring a comprehensive understanding of their impact. Section V presents the fit results for the valence distribution functions, their evolution, corresponding errors, and our findings regarding $\alpha_{s}(M_Z^2)$ at the NNLO. We also compare these results with other theoretical outcomes to establish their significance and reliability. Finally, we conclude our study by discussing the results in Section VI.

\section{DIS structure functions and parton distributions}
The present flavor non-singlet analysis is based on four complementary data sets: the non-singlet structure function $xF_3(x,Q^2)$, as well as $F_2^p(x,Q^2)$ and $F_2^d(x,Q^2)$ structure functions in the valence-quark region $x \geq 0.3$, and the combination of $F_2^{p,d}(x,Q^2)$ is employed to determine $F_2^{NS}(x,Q^2)=2[F_2^p(x,Q^2)-F_2^d(x,Q^2)]$.
Therefore, we proceed to introduce these structure functions in this section. Theoretical results of these structure functions, including anomalous dimensions, the massless Wilson coefficients, and the massive Wilson coefficients up to three-loop orders, are described in Refs.~\cite{Moch:2002sn, Moch:2004pa, Vogt:2004mw, Blumlein:2021enk, Politzer:1974fr, vanNeerven:1999ca, Vermaseren:2005qc, vanNeerven:1991nn, Zijlstra:1991qc, Zijlstra:1992kj, Blumlein:2022gpp, Ablinger:2017err, Ablinger:2012ej}. Also, one can refer to a good review of Ref.~\cite{Blumlein:2023aso} and references therein.

The charged-current (CC) deep inelastic neutrino (antineutrino)-nucleon scattering differential cross sections at the leading order of the running coupling constant $\alpha_s$ can be described in terms of structure functions related to parton distributions \cite{Eisele:1986uz,Diemoz:1986kt}. 
In the quark parton model (QPM) framework, the structure of neutrinos and antineutrinos interacting with nucleons is described at the leading order (LO) regarding their connection to the valence PDFs. Within the QPM, the averaged nucleon structure for neutrinos and antineutrinos is exclusively governed by the valence quark distributions. 
Investigating the DIS neutrino (antineutrino)-nucleus $xF_3$ structure function requires considering the nuclear effect due to neutrino detection involving heavy nuclear targets \cite{deFlorian:2011fp}. Various neutrino experiments employ nuclear targets, such as CCFR, NuTeV, CDHSW (using an iron target), and CHORUS (using a lead target). Nuclear PDFs are crucial in obtaining the average of the neutrino and antineutrino nucleus structure functions. 
In Ref.~\cite{Schienbein:2007fs}, a study on the nuclear effects in CC neutrino-iron DIS data is conducted to determine the iron PDFs. Furthermore, Ref.~\cite{Tooran:2019cfz} comprehensively analyzes nuclear effects on heavy neutrino-nucleus scattering. The average of the neutrino and antineutrino nucleus structure functions is given by
\begin{eqnarray}
xF_3(x,Q^2)&=& xu_v(x,Q^2) + xd_v(x,Q^2) ~.
\end{eqnarray}
In the following sections, we will introduce the nuclear effects for neutrino DIS structure functions (see, e.g., \cite{Tooran:2019cfz}, and references therein).

Alternatively, when considering the non-singlet regime and the valence region with $x\geq0.3$ for $F_2^p$ at LO \cite{Blumlein:2006be,Khorramian:2008yh, Khorramian:2009xz}, the combinations of parton densities can be expressed as
\begin{equation}
F_2^{p}(x,Q^2) = \left[ \frac{1}{18}\, xq_{{\rm
NS,}8}^+ +\frac{1}{6}\, xq_{{\rm NS,}3}^+
\right](x,Q^2)+\frac{2}{9} x\Sigma(x,Q^2)~.
\end{equation}
Considering the neglect of sea PDFs in the region $x\geq0.3$, we can deduce the following relations: $xq_{{\rm NS},3}^+ = xu_v - xd_v$ represents the difference between up ($u_v$) and down ($d_v$) valence quarks. $xq_{{\rm NS},8}^+ = xu_v + xd_v$ represents the sum of up and down valence quarks. Additionally, since sea PDFs can be neglected in this region, the term $\Sigma$ represents the sum of up and down valence quarks. Consequently, in the $x$-space, we obtain $x\Sigma(x) = xu_v(x) + xd_v(x)$. These insights allow us to understand the behavior of quarks in the specified region better and facilitate the simplification of the QCD analysis. Consequently, in the $x$-space, we obtain the following:
\begin{eqnarray}
 F_2^{p}(x,Q^2) &= &\left(\frac{5}{18}\, x q_{{\rm NS,}8}^+
 + \frac{1}{6}\, x q_{{\rm NS,}3}^+\right) (x,Q^2)\nonumber \\
 & = &\frac{4}{9}\, x u_v(x,Q^2)+\frac{1}{9}\, x d_v(x,Q^2)~.
\end{eqnarray}

Alternatively, in the specified region, the combinations of parton densities for $F_2^d$ are also given by \cite{Khorramian:2008yh, Khorramian:2009xz}:
\begin{eqnarray}
 F_2^{d}(x,Q^2) & = &\left(\frac{5}{18}\, x q_{{\rm NS,}8}^+\right) (x,Q^2)\nonumber \\
 & = &
 \frac{5}{18}\, x(u_v+d_v)(x,Q^2)~.
\end{eqnarray}
In the region $x\leq 0.3$, for the difference between the proton and deuteron data, we utilize \cite{Khorramian:2008yh, Khorramian:2009xz} 
\begin{eqnarray}
F_2^{NS}(x,Q^2)&\equiv& 2 (F_2^{p}-F_2^{d})(x,Q^2)=\frac{1}{3}\, x q_{{\rm NS,}3}^+(x,Q^2)\nonumber \\
 & = & \frac{1}{3}\,x(u_v-d_v)(x,Q^2)-\frac{2}{3}\, x(\bar{d}-\bar{u})(x,Q^2)~.\nonumber \\
\end{eqnarray}

In the region $x\leq 0.3$, since sea quarks cannot be neglected for smaller values of $x$, we update the expression for $q_{{\rm NS,}3}^+$ as $u_v-d_v+2(\bar{u}-\bar{d})$. Besides parameterizing the valence PDFs, the non-singlet structure function necessitates the inclusion of the distribution $x(\bar{d}-\bar{u})(x,Q^2)$ as an input. It is important to note that directly extracting the $x(\bar{d}-\bar{u})(x,Q^2)$ distribution from deep inelastic scattering (DIS) analysis is challenging; however, this distribution can be determined by utilizing Drell-Yan data. Very recently, strong evidence has emerged for a flavor asymmetry between the $u$ and $d$ sea PDFs in the proton, as observed in deep-inelastic scattering and Drell-Yan experiments by FNAL E906/SeaQuest Collaboration \cite{SeaQuest:2022vwp}. The reported results of the $(\bar d - \bar u)(x, Q^2)$ differences are deduced for the range $0.13 < x < 0.45$. Interestingly, the latest SeaQuest data demonstrate that the $\bar d (x)$ distribution continues to exceed the $\bar u (x)$ distribution even at the highest $x$ value ($x = 0.45$).

\begin{figure*}[h!]
  \begin{center}
  \resizebox{0.69\textwidth}{!}{\includegraphics{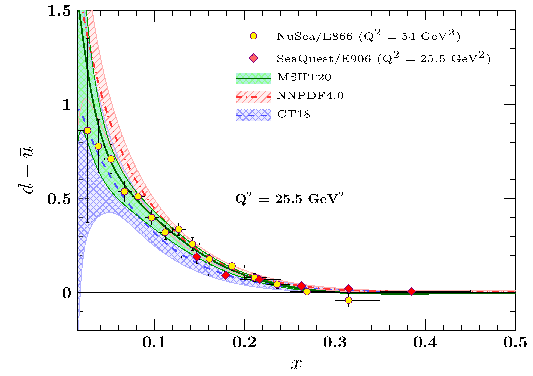}} 
  \caption{  Comparison of recent SeaQuest $\bar{d}(x)-\bar{u}(x)$ results \cite{SeaQuest:2022vwp} and data from NuSea \cite{NuSea:1998kqi,NuSea:2001idv} with the NNLO analysis from MSHT20 \cite{Bailey:2020ooq}, NNPDF4.0 \cite{NNPDF:2021njg}, and CT18 \cite{Hou:2019qau}  as a function of $x$ at Q$^2$=25.5 GeV$^2$.}
  \label{Fig2: dubar Models}
    \end{center}
  \end{figure*}

Figure \ref{Fig2: dubar Models} illustrates recent SeaQuest $\bar{d}(x)-\bar{u}(x)$ results \cite{SeaQuest:2022vwp} in comparison with data from NuSea \cite{NuSea:1998kqi,NuSea:2001idv}. Additionally, the figure displays the modern PDF set results as a function of $x$, which incorporates MSHT20 \cite{Bailey:2020ooq}, NNPDF4.0 \cite{NNPDF:2021njg} and CT18 \cite{Hou:2019qau} at Q$^2$=25.5 GeV$^2$.

In our analysis, we choose to utilize the $x(\bar{d}-\bar{u})(x,Q^2)$ distribution from the MSHT20 \cite{Bailey:2020ooq} set as an input, which incorporates the asymmetry in sea quarks. The MSHT20 PDF set is one of the most recent and modern PDF sets obtained from global fits in a wide range of experimental data, including DIS measurements. While this parametrization has a minor impact on our analysis, it is still essential to assess the influence of this distribution by comparing it with various asymmetrical sea quark distributions derived from other studies. We will discuss the outcomes obtained when incorporating different asymmetry distributions from other analyses. 

 The following section will focus on the discussion of input PDF parametrizations and the experimental measurements corresponding to theories expressed in this section.

\section{Inputs for the Fit: Theoretical and Experimental Aspects}
This section introduces the $xu_v$ and $xd_v$ PDF parametrizations at the input scale of $Q_0^2$. Additionally, we consider $\alpha_s(M_Z^2)$ as another fitting parameter in the present QCD analysis. 
A detailed discussion of various combinations of data sets will be presented for neutrino-nucleon DIS data obtained by CCFR \cite{ccfr:1977}, NuTeV \cite{Tzanov:2005kr}, CHORUS \cite{Onengut:2005kv}, and CDHSW \cite{Berge:1989hr} experiments, as well as charged electron (muon)-nucleon DIS data obtained by BCDMS, NMC, SLAC, H1, and ZEUS experiments \cite{BCDMS:1989ggw, BCDMS:1989qop, BCDMS:1989gtb, Whitlow:1991uw, NewMuon:1996fwh, H1:2000muc, H1:2003xoe, ZEUS:1998agx, ZEUS:2001mhd}. These data sets will be used to determine valence $xu_v$ and $xd_v$ PDFs, as well as $\alpha_s(M_Z^2)$.

\subsection{Valence PDF parametrization}
This section presents the theoretical framework on which our PDF parametrization is based. In the presence of only $xF_3$ DIS data, the determination of the $xd_{v}(x,Q_0^2)$ distribution relies on the $xu_{v}(x,Q_0^2)$ distribution \cite{Tooran:2019cfz,Diemoz:1987xu,Gluck:1989ze,Gluck:1998xa}. Specifically, in the presence of $xF_3$ DIS data, it is not possible to determine $xu_{v}(x,Q_0^2)$ and $xd_{v}(x,Q_0^2)$ separately in a QCD non-singlet analysis. In this case, the $xd_{v}(x,Q_0^2)$ distribution depends on $xu_{v}(x,Q_0^2)$ \cite{Diemoz:1987xu,Gluck:1989ze,Gluck:1998xa}. For example, in pure $xF_3$ QCD analysis \cite{Khorramian:2006wg,Tooran:2019cfz}, it is assumed that $xd_{v}(x,Q_0^2) \propto (1-x)^{b_{d}} xu_{v}(x,Q_0^2)$.

In the presence of only $F_2$ DIS data \cite{Khorramian:2008yh, Blumlein:2006be}, and also both $xF_3$ and $F_2$ DIS data, one can consider a separate parametrization for $xu_{v}(x,Q_0^2)$ and $xd_{v}(x,Q_0^2)$. To proceed with our previous works on $xF_3$ structure function \cite{Khorramian:2006wg,Tooran:2019cfz}, we are interested in linking $u$ and $d$-valence PDF parametrizations. To fulfill this purpose, we utilize the CJ formalism, as mentioned in \cite{Accardi:2016qay, Owens:2012bv}, to provide greater flexibility in determining the $d$-valence PDF, especially at large values of $x$. With this modification, we gain more flexibility in determining the valence $xd_{v}$ in the large $x$, which is important for further studies related to this region in the rest of this paper.
\\
Performing a non-singlet QCD analysis based on $xF_3$ DIS data and adding a world data set for $F_2$ can significantly improve the precision of the valence PDF determination. This approach allows us to present a detailed parametrization of the valence PDF and its correlated errors over a wide range of $x$ and $Q^2$. 

The chosen $u$-valence PDF at the input scale of $Q_0^2$ is expressed as follows
 \begin{eqnarray}
 xu_v(x,Q_0^2)&=&{N_{u_v}}x^{a_{u}}(1-x)^{b_{u}}(1+c_{u}x+d_{u}\sqrt{x})\;, 
 \label{eq:parm1}
\end{eqnarray}
which in the parametrization provided above, we utilize the same parametrization as employed in our previous analysis for $xu_v$ 
\cite{Khorramian:2006wg,Tooran:2019cfz}.

For the $d$-valence PDF, the parametrization is given by
\begin{eqnarray}
 xd(x,Q_0^2)&=& N_{d_v}\left( x^{a_{d}}(1-x)^{b_{d}}(1+c_{d}x+d_{d}\sqrt{x})\right. \nonumber \\
 &&~~~~~~~+ \left.e_d x^{f_d}~xu_v(x,Q_0^2)\right)~.
 \label{eq:parm2}
\end{eqnarray} 
As mentioned, we employ an alternative form for $d$-valence PDF by combining $u$-valence PDF, following the approach outlined in Ref.~\cite{Accardi:2016qay}. This combination permits enhanced flexibility within the large $x$ region. The term $x^{a_{i}}$ control the behavior of valence PDFs in the low $x$ and $(1-x)^{b_{i}}$ control large $x$ regions. In contrast, other polynomial terms play a significant role in intermediate values of $x$. As a result of the modification for $d$-valence PDF parametrization, 
we have $xd(x,Q_0^2)/xu(x,Q_0^2) \rightarrow N_{d_v}e_d~$ in the limit of $x \rightarrow 1$ if $b_d > b_u$.

Additionally, the normalization constants $N_{u}$ and $N_{d}$ can be derived from the other parameters by applying the conservation of the fermion number, which is expressed as
\begin{equation}
\int_0^1u_{v}~dx=2\;,
\end{equation}
\begin{equation}
\int_0^1d_{v}~dx=1\;.
\end{equation}
Using above equations, the normalization constants $N_{u}$ and $N_{d}$ are as follows
\begin{eqnarray}
N_{u}&=& 2 / \big((B(a_{u},1+b_{u}) + c_{u}~B(1+a_{u},1+b_{u}) \nonumber \\&+& d_{u}~B(1/2+a_{u},1+b_{u})\big),\\
N_{d} &=& 1 / \bigg(B\left(a_d, 1 + b_d\right) + c_d~B\left(1 + a_d, 1 + b_d\right) \nonumber \\
&+& d_d~B\left(1/2 + a_d, 1 + b_d\right) \nonumber \\
&+& N_{u}~{e_d} ~\Big(B\left(a_u + {f_d}, 1 + b_u\right) \nonumber \\
&+& c_u~B\left(1 + a_u + {f_d}, 1 + b_u\right) \nonumber \\
&+& d_u~B\left(1/2 + a_u + {f_d}, 1 + b_u\right)\Big)\bigg)~.
\end{eqnarray}

In the above parametrization, $B(a,b)$ represents the Euler $\beta$ function, where the normalization constants $N_{u}$ and $N_{d}$ play a crucial role in determining the unknown parameters through the QCD fitting procedure. They effectively constrain the parameter space and ensure consistency with the conservation of the fermion number.

According to the above parametrization, we may have ten free valence parameters that can be determined through QCD fits. However, in the next section, we will observe that specific parameters need to be fixed after the initial minimization, because the available DIS data may not sufficiently constrain some of the parameters in Eqs.~(\ref{eq:parm1}) and (\ref{eq:parm2}). In particular, the errors associated with some parameters can be relatively large compared to their central values. To address this issue, it is common practice to fix these parameters after the first minimization, as has been done in previous non-singlet QCD analyses of $F_2(x,Q^2)$. Similar procedures have been previously conducted in Refs.~\cite{Khorramian:2009xz,Khorramian:2008yh,Blumlein:2006be}. By fixing these parameters, we ensure stability and avoid overfitting, considering the limitations of the available DIS data and the associated uncertainties in parameter determination.

In general, the coupling constant $\alpha_{\rm s}(M_Z^2)$ can be extracted from global QCD fits to various hadronic processes. However, in this non-singlet QCD analysis, the strong coupling constant at the scale of $M_Z^2$ is considered a free parameter and can be determined using DIS data. The strong coupling constant plays a crucial role, exhibiting a significant correlation with the PDFs. Since the determination of $\alpha_{\rm s}(M_Z^2)$ is connected to uncertainties in other non-singlet PDF parameters, assessing the uncertainty associated with this parameter is essential.

One way to evaluate this uncertainty is by comparing the fitted value of $\alpha_{\rm s}(M_Z^2)$ with the world average which reported as ${\alpha_s(M_Z^2)} = 0.1179 \pm 0.0009$ in Ref.~\cite{ParticleDataGroup:2022pth}. This comparison helps validate the consistency of the obtained value with the experimental measurements and theoretical expectations for the strong coupling constant $\alpha_{\rm s}(M_Z^2)$ in the context of the non-singlet QCD analysis.

In this work, we employed the QCDNUM evolution engine \cite{Botje:2010ay} to determine the $Q^2$ evolution of PDFs and the coupling constant. QCDNUM is a widely used tool in high-energy physics, known for performing precise calculations related to the evolution of PDFs and the running of the strong coupling constant. By utilizing QCDNUM in our analysis, we could accurately account for the $Q^2$ dependence of the PDFs and the corresponding evolution of the strong coupling constant within the framework of QCD.

\subsection{Experimental data sets}
In this section, we employ various experimental data sets for our analysis, which include neutrino-nucleon DIS data from CCFR, NuTeV, CHORUS, and CDHSW experiments, as well as charged electron (muon)-proton and deuteron DIS data from BCDMS, NMC, SLAC, H1, and ZEUS experiments. These diverse data sets offer valuable insights for determining and analyzing the parameters and distributions. By incorporating four different data sets: $xF_3$, $F_2^p$, $F_2^d$, and $F_2^{NS}$, a more comprehensive understanding of the nucleon structure and the behavior of valence quarks can be obtained. This expanded approach allows for a more accurate description of the experimental data and enhances the precision of the extraction process for valence quark densities and the strong coupling constant.

In this analysis, we include the DIS neutrino (antineutrino) $xF_3$ measurements from CCFR [$30\leq E_\nu$(GeV)$ \leq 360$] with an iron target \cite{ccfr:1977}, NuTeV [$30\leq E_\nu$(GeV)$ \leq 500$] with an iron target \cite{Tzanov:2005kr}, CHORUS [$10\leq E_\nu$(GeV)$ \leq 200$] with a lead target \cite{Onengut:2005kv}, and CDHSW [$20\leq E_\nu$(GeV)$ \leq 212$] with an iron target \cite{Berge:1989hr}. Nevertheless, neutrino-nucleon $xF_3$ experiments have employed targets with high atomic number (Z) and mass number (A), such as iron or lead. Therefore, it is worth considering the nuclear corrections \cite{deFlorian:2011fp} to have a good accuracy.

In addition, for the current QCD analysis, we include the structure function data obtained from charged electron (muon)-proton and deuteron DIS experimental data, such as BCDMS \cite{BCDMS:1989ggw, BCDMS:1989qop, BCDMS:1989gtb}, SLAC \cite{Whitlow:1991uw}, NMC \cite{NewMuon:1996fwh}, H1 \cite{H1:2000muc, H1:2003xoe}, and ZEUS \cite{ZEUS:1998agx, ZEUS:2001mhd}. These experiments mentioned above provide a reliable and accurate foundation for determining valence-quark densities and $\alpha_s(M_Z^2)$, thus enhancing the precision of our analysis. Also, utilizing deuteron data in the high $x$ region requires nuclear correction, which we will describe in the next section.

Therefore, all the data sets used in this analysis include $xF_3 (x,Q^2)$ data, $F_2^p (x,Q^2)$, $F_2^d (x,Q^2)$ in the valence quark region $x \geq 0.3$, and finally $F^{NS}_2 = 2(F^p_2 - F^d_2)$ in the region $x < 0.3$. We approximate the PDFs by considering only pure valence quarks within the valence quark region. 

In this analysis, two different fits at the NNLO with the names of $\mathtt{SK24[QCD]}$ and $\mathtt{SK24[QCD+TMC+HT]}$ are introduced below: 

\begin{itemize}
	
\item \textbf{$\mathtt{SK24[QCD]}$}:
In the first fit, only data with $Q^2 \geq 4 \text{ GeV}^2$ are included, and a cut in the hadronic mass of $W^2 \geq 12.5 \text{ GeV}^2$ is applied to eliminate higher twist effects from the data samples effectively. 

\item \textbf{$\mathtt{SK24[QCD+TMC+HT]}$}: In the second fit, we apply the same cuts as in \textbf{$\mathtt{SK24[QCD]}$}, with the exception that we impose $W^2 \geq 4\text{ GeV}^2$. In this scenario, it is necessary to consider target mass corrections and higher-twist effects.

\end{itemize}
 As we will see, the imposition of cuts on $W^2$ in the datasets used in this analysis substantially enhances the number of available data points, contributing to a more effective constraint on valence PDFs. In both fits, some unknown parameters exist for valence PDFs and $\alpha_s(M^2_Z)$. 
 The MINUIT \cite{James:1975dr} program is employed for the minimization in the fitting procedure.

After applying $Q^2 \geq 4 \text{ GeV}^2$ and $W^2 \geq 12.5 \text{ GeV}^2$ cuts, we are left with 1045 data points: 283 data points for $xF_3$, 322 data points for $F^p_2$, 232 data points for $F^d_2$, and 208 data points for $F^{NS}_2$.
We should note that three additional cuts need to be applied for the neutrino-nucleon and lepton-nucleon DIS data. A $x > 0.4$ cut is necessary for CCFR data due to a discrepancy between CCFR and NuTeV in this region \cite{Tooran:2019cfz}. Additionally, to exclude a region with significant correlated systematic errors, a $y_\mu > 0.3$ cut is necessary for the BCDMS data \cite{Blumlein:1992we}. Also, the NMC data requires an extra cut of $Q^2 \geq 8 \ \text{GeV}^2$. For more information on these additional cuts, please see Refs.~\cite{Khorramian:2006wg,Tooran:2019cfz,Blumlein:1992we}.
After these additional cuts, the total number of DIS data points has been reduced from 1045 to 814. This includes 263 data points for $xF_3$, 227 data points for $F^p_2$, 159 data points for $F^d_2$, and 165 data points for $F^{NS}_2$. By considering $W^2 \geq 4\text{ GeV}^2$ on all data, the total increases to 1395, including 287 data points for $xF_3$, 506 data points for $F^p_2$, 437 data points for $F^d_2$, and 165 data points for $F^{NS}_2$.

Table \ref{tab:deltachi2fitted} presents different combinations of subsets of non-singlet $xF_3$ and $F_2$ data, along with their corresponding $x$ and $Q^2$ ranges, used in the current analyses. The fourth and fifth columns show the number of individual data points for each data set with $W^2\geq$ 12.5 GeV$^2$ and $W^2\geq$ 4 GeV$^2$ cuts applied to the data. The sixth and seventh columns ($\mathtt{SK24[QCD]}$) and the eighth and ninth ($\mathtt{SK24[QCD+TMC+HT]}$) display the $\chi^2$ and $\chi^2/$\#Data values for each fit applied on corresponding cuts. The $\chi^2$ and total $\chi^2/d.o.f.$ are also shown in this table.

\begin{figure*}[h!]
\begin{center}
\resizebox{0.45\textwidth}{!}{\includegraphics{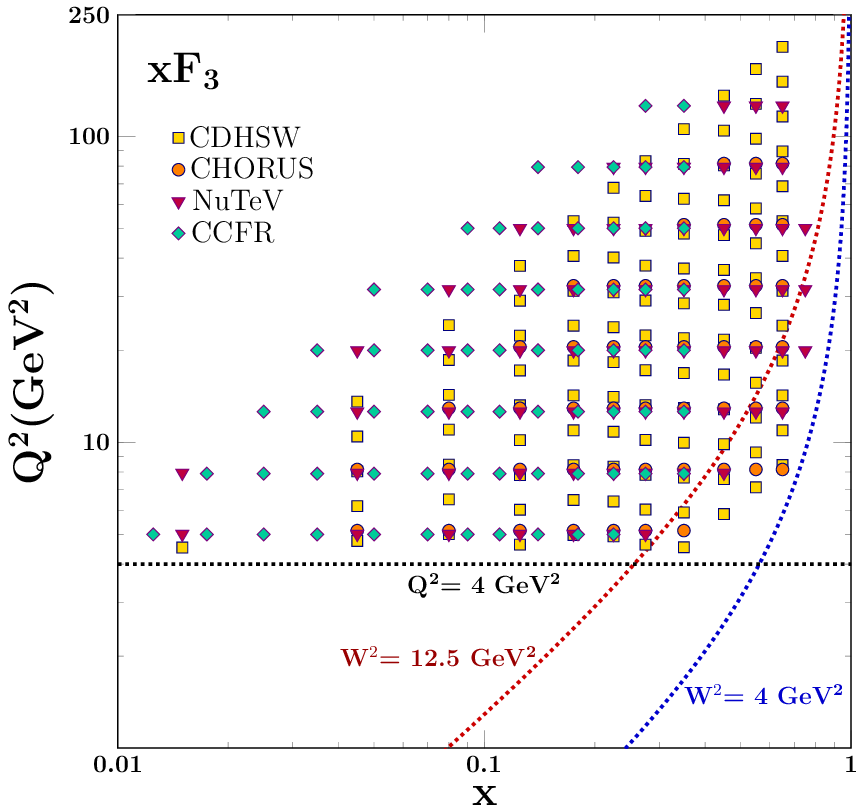}} 
\resizebox{0.45\textwidth}{!}{\includegraphics{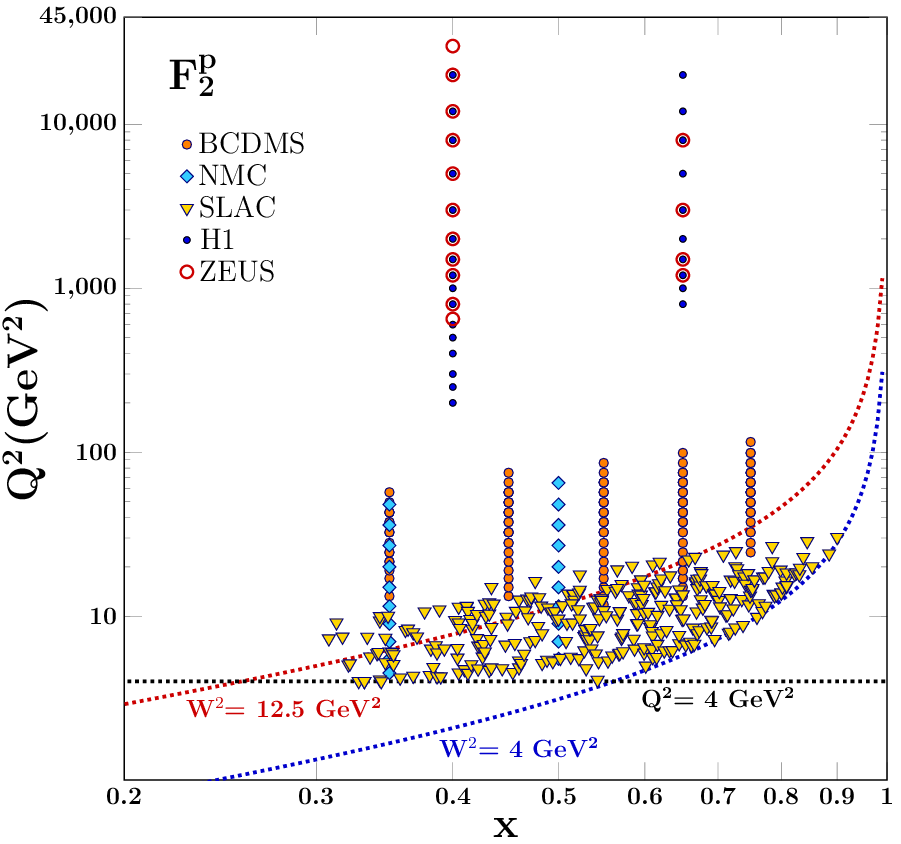}}

\resizebox{0.45\textwidth}{!}{\includegraphics{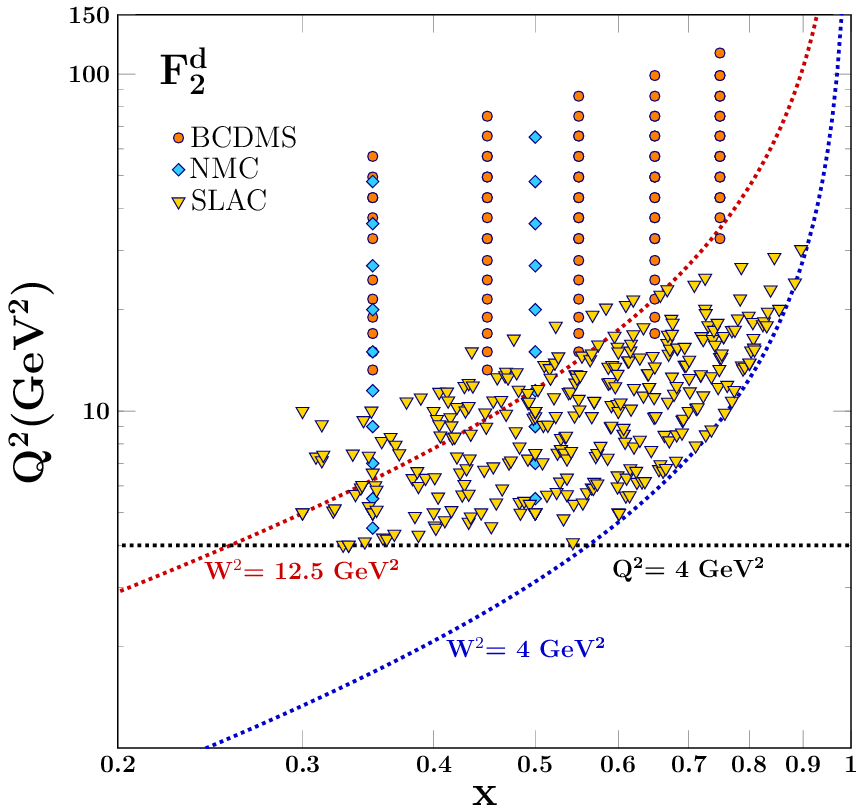}} 
\resizebox{0.45\textwidth}{!}{\includegraphics{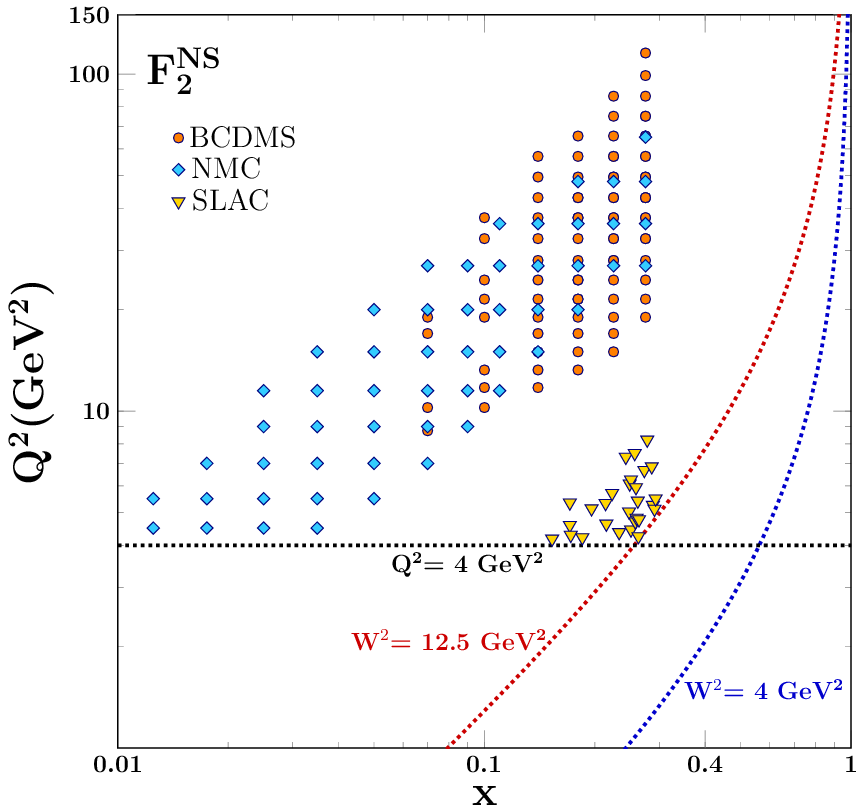}} 
\caption{The $x$ and $Q^2$ plane of various experiments of DIS lepton-nucleon data for $xF_3(x,Q^2)$, $F_2^p(x,Q^2)$, $F_2^d(x,Q^2)$ and $F_2^{NS}(x,Q^2)$. The dotted lines represent the kinematic cuts applied in the analysis, which require $Q^2\geq 4$ GeV$^2$, $W^2\geq 12.5$ GeV$^2$ and $W^2\geq 4$ GeV$^2$. Data points above the $W^2=12.5$ GeV$^2$ line are used for QCD fits ($\mathtt{SK24[QCD]}$), while data points above the $W^2=4$ GeV$^2$ line are used for QCD+TMC+HT fits ($\mathtt{SK24[QCD+TMC+HT]}$) in the current analysis.}
\label{Fig3:Kinematic}
  \end{center}
\end{figure*}

In Figure \ref{Fig3:Kinematic}, we present the experimental data utilized in this analysis, including neutrino-nucleon DIS data extracted from CCFR, NuTeV, CHORUS, and CDHSW experiments, as well as the electron (muon)-proton and deuteron DIS experimental data obtained from BCDMS, NMC, SLAC, H1, and ZEUS experiments, plotted in the $x$ and $Q^2$ plane. The data points are selected by imposing cuts on $W^2$ ($W^2\geq$ 12.5 GeV$^2$ and $W^2\geq$ 4 GeV$^2$) and $Q^2\geq$ 4 GeV$^2$. As previously mentioned, a discrepancy exists between NuTeV and CCFR at $x>0.4$, and consequently, the CCFR data in this region is excluded from the analysis. Please see Ref. \cite{Tooran:2019cfz} for more information and details. As seen from Figure \ref{Fig3:Kinematic}, a significant number of data points for both $F_2^p$ and $F_2^d$ are found within the range of $4 \leq W^2 \leq 12.5, \text{GeV}^2$. However, there is no available data for $F_2^{NS}$ within this specific region. 

In the upcoming section, we will introduce corrections related to the data selection discussed earlier and the theories explained in Sec. II.

\section{QCD Corrections}
In this section, our main focus is on QCD corrections, specifically emphasizing nuclear effects, target mass corrections, and higher twist effects.
Through a detailed investigation of these components, we aim to study their impact on the present non-singlet QCD analysis of DIS structure function data.

\subsection{Deuteron nuclear correction}
 Implementing deuteron nuclear corrections becomes challenging in the high $x$ region due to the essential requirement for deuteron wave function (DWF) and nuclear off-shell function (OSF), making it impossible to neglect this effect \cite{Alekhin:2017fpf}. In the present paper, due to the differences between the PDFs in deuterons and free nucleons, and also the use of deuteron DIS data in very high $x$ region ($x \gtrsim 0.7$), we are required to implement deuteron nuclear correction concerning large $x$ \cite{Accardi:2016qay, Kovarik:2015cma}. Therefore, we follow the approach investigated by S. A. Kulagin and R. Petti in Ref.~\cite{Kulagin:2004ie}. In this instruction, the deuteron structure function becomes \cite{Alekhin:2017fpf, Kulagin:2004ie}
 \begin{eqnarray}
  \label{eq:IA}
  && \gamma^2 F_2^{d}(x,Q^2) = \int \frac{\mathrm{d}^3\bm p}{(2\pi)^3} \left|\Psi_D(\bm p)\right|^2 \notag \\
  && \left(1+\frac{\gamma p_z}{M}\right) \left({\gamma'}^2 +\frac{6{x'}^2 \bm{p}_\perp^2}{Q^2} \right) F_2^N(x',Q^2,p^2)~.
  \end{eqnarray}
In the above equation $\gamma^2=1+4x^2 M^2/Q^2$ and ${\gamma'}^2=1+4{x'}^2 p^2/Q^2$ and also $\Psi_D(\bm p)$ is a deuteron wave function which is depends on the momentum of bound nucleon $\boldsymbol{p}=\left(\boldsymbol{p}_{\perp}, p_z\right)$ and can archive from various analyses such as AV18 \cite{Veerasamy:2011ak}, WJC1 \cite{Gross:2008ps} and WJC2 \cite{Gross:2010qm}.
  
Mass of bound nucleon can be written as $M = \frac{1}{2} (M_p + M_n)$ and Bjorken scaling as $x' = x /\left[1+\left(\varepsilon+\gamma p_z\right) / M\right]$, in this relation $\varepsilon=\varepsilon_D-\bm p^2/(2M)$ where deuteron binding energy define as $\varepsilon_D=M_D-2M$. $F_2^{\smash{N}}=\tfrac12(F_2^p+F_2^n)$ is the structure function of the bound nucleon which its $p^2$-dependence comes from off-shell dependence of the LT structure function
 \begin{align}
  F_2^{\mathrm{LT}}\left(x, Q^2, p^2\right) & =F_2^{\mathrm{LT}}\left(x, Q^2\right)\left[1+\delta f\left(x, Q^2\right) v\right]~,
 \end{align}
 here $v= \left(p^2 - M^2\right) / M^2$ is nucleon virtuality. For the off-shell effect $\delta f (x)$ we consider a phenomenological model which is taken by Refs.~\cite{Accardi:2016qay, Kulagin:2004ie} 
 \begin{align}
  \label{eq:delf}
  \delta f^N=C_N\left(x-x_0\right)\left(x-x_1\right)\left(1+x_0-x\right)~,
  \end{align}
where $C_N$ and $x_0$ are free parameters participating in fit procedure and $x_1$ claculated analytically to conserve number of valence quarks. For more details of this instruction, please see Refs.~\cite{Kulagin:2004ie, Alekhin:2017fpf}, and also for the application of this instruction at the partonic level, please look at Ref.~\cite{Accardi:2016qay} 
\\
In the present paper, we consider WJC2 \cite{Gross:2010qm} for the deuteron wave function, and we follow the same treatment for the nuclei structure functions as our previous analysis in Ref.~\cite{Tooran:2019cfz}.

\subsection{Target mass corrections}
 
In our non-singlet analysis, another correction that needs to be included is TMCs. The effects of target masses were first discussed in the context of operator-product expansion (OPE) in Ref.~\cite{Georgi:1976ve}. Indeed, to ensure the desired accuracy of theory calculations, it becomes essential to account TMCs in the present analyses of non-singlet structure functions. One crucial aspect that must be considered is the target mass correction in twist $\tau=2$ structure functions for nuclei. Based on the data sets mentioned in the previous section and the wide range of kinematics covered in the present analysis, it is necessary to consider the impact of TMCs on unpolarized proton and nuclear targets. Very recently, a detailed derivation of the formalism of TMCs to structure functions in DIS has been presented in Ref.~\cite{Ruiz:2023ozv}.

An accurate implication of TMCs, which is subleading $1/Q^2$ correction on structure function in the leading twist approximation, is vital to reliable PDFs determination and also comprehensive data interpretation. The impact of this correction is considerable in large $x$ and moderate $Q^2$ regions. Higher values of $Q^2$ decrease the effect of TMCs due to elimination of the mass term in Eq.~\eqref{eq:RA} and subsequently, $\xi_A \rightarrow x_A$ in Eq.~\eqref{eq:Xi}. In terms of this notation, $x_A$ is Bjorken scaling variable, also Nachtmann variable $\xi_A$ and quantity of $r_A$ define as
\begin{equation}
 r_A\ =\ \sqrt{1 + 4 x_A^2 M_A^2/Q^2}\, ,
 \label{eq:RA}
\end{equation}
\begin{equation}
\xi_A\ =\ \frac{2x_A }{1 + r_A}\ \equiv R_M x_A, ~~~~R_M=\frac{2}{1 + r_A}~.
\label{eq:Xi}
\end{equation}
In the above, $R_M$ denotes the factor dependent on the target mass, connecting the Bjorken scaling variable $x_A$ to the Nachtmann scaling variable $\xi_A$.

According to the master formulas for twist $\tau=2$ target mass corrections to structure functions for nuclei, we have the $Q^2$ dependence of TMCs to structure functions as 
\begin{eqnarray}
\label{eq:TMCa} 
{F}_{2}^{A,TMC}(x_A) &=& \left(\frac{x_A^{2}}{\xi_A^{2}r_A^{3}} \right) {F}_{2}^{A,(0)}(\xi_A)\nonumber \\
 &&+\left(\frac{6 M_A^2 x_A^{3}}{Q^2 r_A^{4}}\right) \, {h}_{2}^A(\xi_A)\nonumber \\
 &&+\left(\frac{12 M_A^4x_A^{4}}{Q^4 r_A^{5}}\right) \, {g}_{2}^A(\xi_A)\, ,
 \\
 \label{eq:TMCb}
x_A{F}_{3}^{A,TMC}(x_A) &= & \left(\frac{x_A^2}{\xi_A r_A^{2}}\right) {F}_{3}^{A,(0)}(\xi_A)
\nonumber \\
 &&+\left(\frac{2 M_A^2x_A^{3}}{Q^2 r_A^{3}}\right) \, {h}_{3}^A(\xi_A)\, . 
\end{eqnarray}
Here, the $Q^2$ dependence of the structure functions is omitted for brevity. In the present analysis, $A = \{1, 2,56, 197\}$ stand for mass number related to $F_2^{p, NS}$, $F_2^{d}$, $xF_3^{Fe}$ and $xF_3^{Pb}$, respectively. Also, ${F}_{2, 3}^{A,(0)}$ are massless nuclei structure functions when $M^2 / Q^2 \rightarrow 0$ and one can find coefficients of this structure functions, ${h}_{2, 3}^A$ and ${g}_{2}^A$ in Refs.~\cite{Kretzer:2003iu}.

In Eqs.~(\ref{eq:TMCa}, \ref{eq:TMCb}), the functions ${h}_{2}^A(\xi_A)$, ${g}_{2}^A(\xi_A)$ and ${h}_{3}^A(\xi_A)$ are expressed as integrals
\begin{eqnarray}
\label{eq:TMC2a}
{h}_{2}^A(\xi_A,Q^{2}) & = \int_{\xi_A}^{1}du_A\ \frac{{F}_{2}^{A,(0)}(u_A,Q^{2})}{u_A^{2}},
\\
{g}_{2}^A(\xi_A,Q^{2}) & = \int_{\xi_A}^{1}du_A\ 
{h}_{2}^A(u_A,Q^{2})~,\\
\label{eq:TMC2b}
{h}_{3}^A(\xi_A,Q^{2}) & = \int_{\xi_A}^{1}du_A\ \frac{{F}_{3}^{A,(0)}(u_A,Q^{2})}{u_A}~.
\end{eqnarray}
By incorporating Eqs.~(\ref{eq:TMCa}, \ref{eq:TMCb}) and also (\ref{eq:TMC2a} $\!\!-\!\!$ \ref{eq:TMC2b}), we can consider the TMC effects at twist $\tau=2$ structure functions for nucleons and nuclei in the current analysis. 

The accuracy of the calculation of the TMCs has a significant effect on our further study of higher twist correction, this impact can be seen by the values and uncertainties of its parameters. For reliable implication and increasing precision of TMCs calculation like integration and interpolation, we used the MBUTIL package released by QCDNUM \cite{Botje:2010ay}. This FORTRAN package involves public libraries CERNLIB and NETLIB and some privately developed routines.

In the following subsection, we discuss higher twists and their application to target mass corrections, involving them as free parameters directly in the fit.

\subsection{Higher twist effects}

The final correction considered in the present analysis involves HT effects. 
These effects significantly contribute to improving the accuracy of the present analysis. By considering HT effects, a more complete and detailed understanding of the data can be achieved. Continuing with the analysis, we further investigate the influence of HT effects on the current QCD analysis of the $xF_3$ and $F_2$ structure function measurements. By explicitly accounting HT effects, we aim to gain insights into the additional nonperturbative contributions to the structure functions. 

In the standard analysis of DIS structure function data, it is essential to incorporate appropriate cuts on the invariant-mass squared $W^2=Q^2(1/x-1)+m_N^2$ and the virtual photon $Q^2$ at the NNLO. Choosing an appropriate $W^2$ cut value for the structure function data is crucial to disregard the influence of nonperturbative effects. By applying these cuts, the analysis can focus primarily on the perturbative QCD aspects of the data.

To address the HT effects in the data, we initially applied standard cuts on $Q^2$ and invariant-mass squared $W^2$. Specifically, we implemented cuts $Q^2\geq4$ GeV$^2$ and $W^2\geq$12.5 GeV$^2$ to exclude the influence of HT effects from the data. By employing these cuts, our objective was to minimize nonperturbative contributions and concentrate primarily on the perturbative QCD aspects of the analysis. Following the application of these cuts, we proceeded with the extraction of the unknown parameters through QCD fits on the data. Through this fit, we aimed to determine the best-fit values of the parameters and gain a more precise understanding of the QCD dynamics underlying the structure function measurements.

In the subsequent step, we investigated the impact of the HT contribution by utilizing all available data within the $Q^2\geq4$ GeV$^2$ and $W^2\geq4$ GeV$^2$ region. This approach allowed us to incorporate data spanning the DIS region into our QCD fits, enabling a comprehensive analysis of the HT effects on the structure function measurements. In this context, notable efforts have been made in previous works such as Refs.~\cite{Kataev:1997vv,Blumlein:2008kz,Sidorov:1996wb,Sidorov:1996if,Virchaux:1991jc} to address the HT contributions and investigate their implications.

To account for the HT contribution, the average of DIS structure functions can be expressed as a combination of the perturbative and nonperturbative contributions
\begin{eqnarray}
 F_i(x,Q^2) = F_i{\mathtt{[QCD+TMC]}}(x,Q^2)\left(1+\frac{C^i_{HT}(x)}{Q^2}\right) ~.
 \label{eq:HT}
\end{eqnarray}
Here, the $Q^2$ dependence of the first term is obtained through perturbative QCD and TMC, and the coefficient function for 
higher twists is parameterized by a polynomial function, expressed as \cite{Accardi:2016qay, Owens:2012bv}
\begin{eqnarray}
C^i_{HT}(x)= h_{i1} x^{h_{i2}} (1 + h_{i3} x) ~,
\label{eq:HTform}
\end{eqnarray}
where $i=1,2,3$ present the HT effects for the $F_2^{p}$, $F_2^{d}$, and $xF_3$ structure functions, respectively. Through a comprehensive fitting procedure using experimental data, the unknown parameters of $h_{ij}$ with $j=1,2,3$ and their corresponding uncertainties for the function $C^i_{HT}(x)$ can be simultaneously extracted along with other unknown parameters, including those associated with the valence PDFs and the strong coupling constant.

 As we discussed, the fitting process allows us to incorporate the modifications arising from nuclear effects, target mass corrections, and higher twist effects in our analysis. These modifications are essential for achieving a more accurate description of the data.

\section{Results}
A comprehensive non-singlet QCD analysis is presented in this work, focusing on the structure functions of both neutrino-nucleon and electron-nucleon deep inelastic scattering. The analysis encompasses the charged current neutrino-nucleon DIS data, corresponding NNLO approximations, accounting for nuclear and higher twist corrections. Therefore, the world data for lepton scattering is employed for the NNLO analysis of the DIS structure functions $xF_3 (x,Q^2)$, $F_2^{p} (x,Q^2)$, $F_2^{d} (x,Q^2)$, and $F_2^{NS} (x,Q^2)$. This analysis is performed to determine the parameters of the valence PDFs together 
$\alpha_s(M_Z^2)$ with the correlated errors.

This analysis performs two separate fits, considering the cut studies in both $Q^2$ and $W^2$. Remarkably, it has been observed that power corrections are essentially negligible in the kinematic region Q$^2\geq 4$ GeV$^2$ and $W^2 > 12.5$ GeV$^2$. Consequently, we have extended these cuts to the entire data set, considering HT corrections. Additionally, we have considered the TMCs by unfolding them for all the data and nuclear effects for lepton-nucleon structure functions.

\begin{figure*}[h!]

\includegraphics[width=0.6\textwidth]{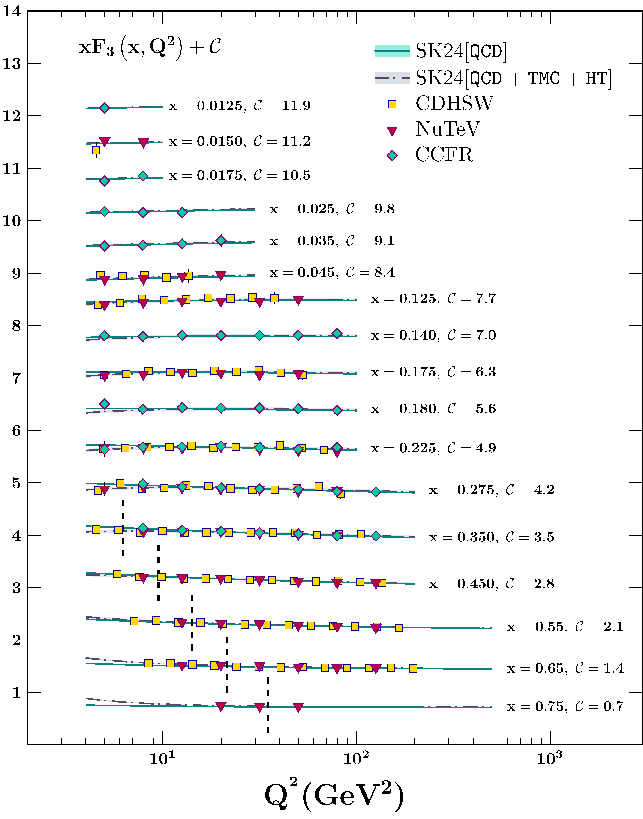}
  
\caption{The comparison of structure function $xF_3$ (Iron target) obtained from the $\mathtt{SK24[QCD]}$ and $\mathtt{SK24[QCD+TMC+HT]}$ fits  as a function of $Q^2$ in the various $x$, at the NNLO approximation. The vertical dashed line indicates the regions with W$^2$ $\geq$ 12.5$~$GeV$^2$.}
\label{fig:All-In-One-xF3Fe}
\end{figure*}

Figures \ref{fig:All-In-One-xF3Fe} $\!\!-\!\!$ \ref{fig:All-In-One-F2NS} display the comparison of $xF_3 (x,Q^2)$ and $F_2(x,Q^2)$ structure functions with and without TMC and HT obtained from the different fits at the NNLO as a function of $Q^2$ in various $x$ in the valence quark region. The vertical dashed line indicates the regions with W$^2$ $\geq$ 12.5$~$GeV$^2$.

By conducting this analysis, we extract the valence $xu_v$ and $xd_v$ PDFs, within a wide range of $x$ and $Q^2$. The parametrization of these PDFs is accompanied by their respective correlated errors, providing a comprehensive understanding of their uncertainties.
Table \ref{tab:parm} presents a summary of the fit results for the parameters of $xu_v(x,Q_0^2)$ and $xd_v(x,Q_0^2)$ PDFs, and $\alpha_s(M_Z^2)$ at the NNLO. Also in our analysis, we determine off-shell function $\delta f (x)$, phenomenologically, from Eq.~\ref{eq:delf} during the fit process. Our result for off-shell function parameters satisfies the constraints $0 < x_1 < x_0 < 1$ and $(1 + x_0) > 1$ as mentioned in Ref.~\cite{Kulagin:2004ie}, and confirms the result of this analysis in the shape of the off-shell function. Additionally, the higher twist parameter values and their uncertainties have been reported in this table.   

In all fits, the covariance matrix is verified to be positive definite. Table \ref{tab:QCDcov} represents the values of the covariance matrix for $\mathtt{SK24[QCD]}$ fit, as an example, corresponding to each free parameter. The previous non-singlet analyses in the presence of only $F_2$ DIS data \cite{Blumlein:2006be, Khorramian:2008yh} used two free parameters for each valence PDF, and also, in our previous research in the presence of only $xF_3$ DIS data \cite{Tooran:2019cfz, Khorramian:2006wg} just one free parameter used for $xd_v$ distribution. In the present analysis, due to the combination of $xF_3$ and $F_2$ DIS data and also the parametrization formalism mentioned in Eqs.~(\ref{eq:parm1}, \ref{eq:parm2}), we gain more freedom in valence PDF distributions by using five and six free parameters for SK24[QCD] and SK24[QCD + TMC + HT] fit procedures, respectively. Additional flexibility is achieved from the modified form of $xd_{v}(x,Q^2)$ at the input scale of $Q_0^2$. 

Due to the effect of deuteron nuclear correction on $xd_v$, especially in large $x$ region, we consider different treatments regarding the form of parameterization in $\mathtt{SK24[QCD]}$ and $\mathtt{SK24[QCD+TMC+HT]}$. In the case of $\mathtt{SK24[QCD]}$, the parameter $d_u$ is considered free in the first minimization while $e_d$ and $f_d$ are fixed on zero. That means we consider separate parameterization formalism for each valence PDF in $\mathtt{SK24[QCD]}$ fit procedure. Otherwise, a rise in the $xd_v$ PDF shape in large $x$ region ($x \gtrsim 0.7$) will occur. On the other hand, in the case of $\mathtt{SK24[QCD+TMC+HT]}$, selecting $d_u$ as a free parameter results in a noticeable rise in the $xd_v$ PDF shape in the intermediate $x$ region, influenced by the effective area of this parameter and the dependency of $xd_v$ on $xu_v$ in the parametrization space. We choose $d_u=0$ in this fit to maintain consistency, like CJ PDF analysis \cite{Accardi:2016qay, Owens:2012bv}.

We determined the strong coupling constant, $\alpha_{s}(M_Z^2)$, at the NNLO fits as $0.1154 \pm 0.0009$ and $0.1149 \pm 0.0014$ in the case of $\mathtt{SK24[QCD]}$ and $\mathtt{SK24[QCD+TMC+HT]}$ fits, respectively. The world and DIS averages for ${\alpha_s(M_Z^2)}$ is reported as ${\alpha_s(M_Z^2)} = 0.1179 \pm 0.0009$ and ${\alpha_s(M_Z^2)} = 0.1162 \pm 0.0020$ in Ref.~\cite{ParticleDataGroup:2022pth}. 

 Furthermore, we compare our valence-quark densities results and their uncertainties with predictions obtained using alternative PDF sets from different research groups. Figure~\ref{fig:QCD-HT PDFs} displays the valence $xu_v$ and $xd_v$ PDF comparison between present $\mathtt{SK24[QCD]}$ and $\mathtt{SK24[QCD+TMC+HT]}$ analyses, respectively, at Q$^2$ = 4, 5, 10, 100, 1000 and 5000 GeV$^2$. Also, the middle and right panels of Figure~\ref{fig:QCD-HT PDFs} show the relative uncertainties $\delta xu_v(x,Q^2)/xu_v(x,Q^2)$ and $\delta xd_v(x,Q^2)/xd_v(x,Q^2)$, respectively. 
 
 Figure~\ref{fig:PDFs-xuv-xdv} illustrate our fit results to present the evolution of the valence quark densities $xu_v(x,Q^2)$ and also $xd_v(x,Q^2)$ from $Q^2 = 4 \ \mathrm{GeV}^2$ to $Q^2 = 5000 \ \mathrm{GeV}^2$ in the region $x \in [0.0001, 1]$ at NNLO with correlated errors. Error propagation is carried out using the evolution equations to calculate the associated error bands. First and third columns of these figures compare our results with other NNLO analyses like the results of CT18 \cite{Hou:2019qau} and very recent NNPDF4.0 \cite{NNPDF:2021njg} and MSHT20 \cite{Bailey:2020ooq} analysis. Note that, in the extracted results, we compare combined non-singlet/singlet analyses, and deviations for valence PDFs in smaller values of $x$ may arise due to different assumptions regarding the sea PDFs. Also, as the value of $Q^2$ increases, the distributions tend to flatten at large values of $x$ and rise at low values. Notably, these illustrations include two ratios for comparison with other models. The second and fourth columns of Figure~\ref{fig:PDFs-xuv-xdv} depict log plots for the ratios of $xu_v(x,Q^2)/xu_v(x,Q^2){\text{ref}}$ and $xd_v(x,Q^2)/xd_v(x,Q^2){\text{ref}}$, respectively, with respect to $\mathtt{SK24[QCD]}$. 
 
Figures~\ref{fig:xuv-err} and \ref{fig:xdv-err} display linear plots for relative uncertainties $\delta xu_v(x,Q^2)/xu_v(x,Q^2)$ and $\delta xd_v(x,Q^2)/xd_v(x,Q^2)$, respectively for both $\mathtt{SK24[QCD]}$ and $\mathtt{SK24[QCD+TMC+HT]}$ fit. In these figures, we also compare our results with other NNLO analyses, such as the results of CT18 \cite{Hou:2019qau}, as well as very recent NNPDF4.0 \cite{NNPDF:2021njg} and MSHT20 \cite{Bailey:2020ooq} analyses.

 Figure~\ref{Fig:AlphaS} present the comparison of our extracted value of $\alpha_{s}(M_Z^2)$, at the NNLO with other results obtained from A02~\cite{Alekhin:2002fv}, A06~\cite{Alekhin:2006zm}, MRST03~\cite{Martin:2003tt}, ABMP16~\cite{Alekhin:2017kpj, Alekhin:2018pai}, BBG06~\cite{Blumlein:2006be}, JR14~\cite{Jimenez-Delgado:2014twa}, ABKM09~\cite{Alekhin:2009ni}, ABM11~\cite{Alekhin:2012ig}, KKT~\cite{Khorramian:2009xz}, CT18~\cite{Hou:2019qau}, MSHT20~\cite{Bailey:2020ooq}, NNPDF3.1~\cite{Ball:2018iqk}, NNPDF4.0~\cite{NNPDF:2021njg}, AKP22~\cite{Azizi:2022nqm}, GKA~\cite{Tooran:2019cfz}, and ZEUS~\cite{ZEUS:2023zie}. Furthermore, the world and DIS average values of the strong coupling constant $\alpha_{s}(M_{Z}^{2})$~\cite{ParticleDataGroup:2022pth} are shown. There are potential sources for variation of the coupling constant value $\alpha_{s}(M_{Z}^{2})$. Generally, for this issue, three cases can be mentioned: (a) the theoretical models used in the analyses, such as different treatments for non-perturbative structure function, (b) variation and accuracy of used data, and (c) methodology for implementation of the analysis such as using $x$-space, Mellin-space or usage of the neutral network for fitting procedure. For applying the matching of flavor thresholds at \(Q^2 = m_c^2\) and \(Q^2 = m_b^2\), we chose \(m_c = 1.43 \, \text{GeV}\) and \(m_b = 4.5 \, \text{GeV}\). It is important to note that other results may choose different values for \(m_c\) and \(m_b\). 

It is valuable to examine the influence of individual $xF_3$ data sets on $\alpha_s(M_Z^2)$, identifying one of the sources for variation of this quantity in this analysis comparing to others. To address this aim, we first exclude all four sets of $xF_3$ data. Subsequently, we apply the $\mathtt{SK24[QCD]}$ fitting procedure only on $F_2$ data and determine $\alpha_s(M_Z^2)$ as $0.1142 \pm 0.0009$. This value is consistent with previous analyses of the $F_2$ structure function found in Refs.~\cite{Khorramian:2008yh, Blumlein:2006be}. Expanding this investigation, we reintroduce each set of $xF_3$ data individually. This yields $\alpha_s(M_Z^2)$ values ranging from 0.1152 to 0.1157 for all four possible combinations. Consequently, we conclude that the inclusion of such $xF_3$ data into the $F_2$ structure function dataset leads to an increase in the $\alpha_s(M_Z^2)$ value presented in this analysis compared to our previous study using Bernstein polynomials in Ref.~\cite{Khorramian:2006wg}.

Another method for comparing our NNLO fit results involves calculating moments of $xu_v(x,Q^2)$ and $xd_v(x,Q^2)$ distributions, as well as moments of $xu_v(x,Q^2)-xd_v(x,Q^2)$. The moments of $xq_v$ valence PDFs at the scale of $Q^2$ are as follows:
\begin{equation}
<x^{n-1}>_{q_v}(Q^2)=\int_0^1dx~ x^{n-2} ~xq_v(x,Q^2)~,
\end{equation}
where $n=1,2,3, ...$ corresponds to the first, second, third, etc., Mellin moments, respectively, equivalent to the zeroth, first, second, etc., $x$ moments, and $q_v=u_v,d_v,u_v-d_v$. The first Mellin moments are consistent with the quark number sum rules.

In Table~\ref{tab:LowMom}, we provide the lowest order Mellin moments ($n=2,3,4$) of these valence PDFs at $Q^2=Q_0^2=4$ GeV$^2$ in NNLO and compare them to the corresponding moments obtained from non-singlet GKA19~\cite{Tooran:2019cfz}, KT08~\cite{Khorramian:2008yh}, KT07~\cite{Khorramian:2006wg}, BBG06~\cite{Blumlein:2006be} and combined non-singlet/singlet MSHT20~\cite{Bailey:2020ooq}, NNPDF4.0 \cite{NNPDF:2021njg} and
CT18~\cite{Hou:2019qau} global fits analysis. The $\mathtt{SK24[QCD]}$  moment results of the $xu_v$ PDF demonstrate a closer agreement with BBG06 \cite{Blumlein:2006be}, as well as our KT08 analysis \cite{Khorramian:2008yh} based on the Jacobi polynomial approach. In the case of $xd_v$ moments, our results are closer to NNPDF4.0 \cite{NNPDF:2021njg}, and GKA19~\cite{Tooran:2019cfz}. The $\mathtt{SK24[QCD+TMC+HT]}$ moments of the $xu_v$ PDF shows well compatibility with BBG06~\cite{Blumlein:2006be} and KT08~\cite{Khorramian:2008yh}, also, $xd_v$ PDF moments illustrate close agreement with MSHT20~\cite{Bailey:2020ooq} and KT07~\cite{Khorramian:2006wg}. There are still some deviations and, especially in the case of $xd_v(x, Q^2)$, attributed to variations in data sets, parametrization, high $x$ deuteron nuclear effect, and combinations of non-singlet/singlet QCD analyses.

Impact of the coefficient function for higher twists $C^i_{HT}(x)$ [GeV$^2$] from Eq.~\eqref{eq:HTform} on large $x$ region for the $F_2^{p}$, $F_2^{d}$, and $xF_3$ structure functions corresponding to NNLO fit are presented in Figure~\ref{fig:HTs}. The higher-twist coefficient function $C^i_{HT}(x)$ exhibits growth as $x$ increases. This growth pattern is consistent with the results of the earlier analysis \cite{Blumlein:2006be, Virchaux:1991jc}.

The effects of the TMC and HT corrections are illustrated in Figure~\ref{fig:TMC-Ratio}, where the ratios of the $xF_3$ and $F_2$ structure functions with and without TMC and HT effects are shown, namely $xF_3/xF_3[\mathtt{QCD}]$, $F_2^p/F_2^p[\mathtt{QCD}]$, and $F_2^d/F_2^d[\mathtt{QCD}]$ for Q$^2$ = 4, 10, 25 and 50 GeV$^2$. As seen in this figure, the ratios of structure functions considering TMC and HT, compared to those without these corrections, exceed unity at higher values of $x$. Additionally, with an increase in the value of $Q^2$, the ratio approaches unity. These behaviors are entirely consistent with the theoretical expectations of TMC and HT corrections. 

\section{Conclusions}

We performed a new QCD analysis of the non-singlet world data to extract the valence $xu_v$ and $xd_v$ PDFs, incorporating correlated errors. The article describes a non-singlet QCD analysis using the QCDNUM \cite{Botje:2010ay} open-source framework. Integrating nuclear corrections, TMCs, and HT effects during the fit procedure becomes essential to account for diverse data sets. We have combined these adjustments into our codes, which are not part of the main QCDNUM package.

 In addition to parameterizing the valence PDFs, the non-singlet structure function requires the incorporation of the distribution $x(\bar{d}-\bar{u})(x,Q^2)$ as an input. 
 Although this parametrization has a minor impact on our analysis, it remains crucial to evaluate the influence of this distribution by taking various asymmetrical sea quark distributions derived from other analyses as input. We observed no significant impact when selecting different asymmetry distributions from other studies. Therefore, we used $x(\bar{d}-\bar{u})(x,Q^2)$ from the MSHT20 PDF set due to its novelty and consistency with experimental data.

As mentioned, this analysis is free of assumptions on the gluon to well understanding behaviors of valence PDFs. We investigated the impact of the gluon PDF on all structure functions in our QCD analysis. Our findings indicate that including the gluon contribution does not significantly impact our results. 

Due to different kinematic cuts to the data sets, we applied target mass corrections and higher twist effects. We found that by choosing the same functional form for the $xu_v(x,Q^2)$ and $xd_v(x,Q^2)$ PDFs, a considerable increase is observed in the uncertainty of the coefficient function for higher twists, $C^i_{HT}(x)$, particularly at large values of $x$ for deuteron structure function. So, we considered an alternative form for $d$-valence PDF by combining $u$-valence PDF in a broad range of $x$ and $Q^2$.

In this analysis the higher twist coefficient $C_{HT}(x)$ for $F_2^p$, $F_2^d$, and $xF_3$ is extracted in the region where $W^2\geq$ 4 GeV$^2$. Finally, we determine the strong coupling constant $\alpha_{s}(M_Z^2)$ for our different fits. The obtained results are in good agreement with the world average value and the previous results from different NNLO unpolarized and polarized DIS analyses for $\alpha_{s}(M_Z^2)$.\\

The NNLO grid output for $\mathtt{SK24[QCD]}$ and $\mathtt{SK24[QCD+TMC+HT]}$ analyses are available by contacting the authors via email.
\\

\section{Acknowledgments}
The authors wish to thank the referees for their valuable comments and suggestions, which have significantly enhanced the quality of this paper.
We appreciate F.~Olness for helpful discussion and valuable comments and suggestions. We 
acknowledge A.~Shabanpour for their guidance on computational
programs. We thank F. Gross and W. N. Polyzou for providing the deuteron wave function and their insights during implementation.
A.K. is also grateful to the CERN TH-PH division for the hospitality, where a portion of this work
was performed.

\section*{References}

\begingroup
\squeezetable

\begin{table*}[htp]
 \resizebox{1.0\textwidth}{!}{%
   \begin{tabular}{lclccc@{\hskip 0.1in}c@{\hskip 0.25in}c@{\hskip 0.1in}c}
    \toprule
    \toprule
     & & & \multicolumn{2}{c}{~~\#Data~~~~} & \multicolumn{2}{c}{$\mathtt{SK24[QCD]}$} & \multicolumn{2}{c}{$\mathtt{SK24[QCD+TMC+HT]}$}\\
     \cmidrule(l{2pt}r{2pt}){4-5}	 \cmidrule(l{2pt}r{16pt}){6-7} \cmidrule(l{-2pt}r{0pt}){8-9}
    Data set & $x$ & $Q^2$(GeV$^2$) & {$W^2\geq$ 12.5 GeV$^2$} & {$W^2\geq$ 4 GeV$^2$} & $\chi^2$ & $\chi^2/$\#Data & $\chi^2$ & $\chi^2/$\#Data\\
    \midrule
     $\bf xF_3$\\
    CCFR				    & 0.0125 - 0.35 & 5.00 - 125.9  & {67(87-20)} & 67     & 22.614 &  0.337 & 26.995 &  0.402 \\
    CDHSW				    & 0.0150 - 0.65 & 4.53 - 196.3  & 96          & 107    & 58.750 &  0.611 & 39.246 &  0.366 \\
    NuTeV				    & 0.0150 - 0.75 & 5.01 - 125.89 & 59          & 65     & 85.943 &  1.456 & 64.141 &  0.986 \\
    CHORUS			    & 0.0450 - 0.65 & 5.14 - 81.55  & 41          & 48     & 48.443 &  1.181 & 19.651 &  0.409 \\
    \midrule
    \bf F$\bf_2^{p}$\\
    BCDMS$_{100}$		& 0.35 - 0.75   & 11.75 - 75    & 29 & 39     & 36.045 &  1.242 & 60.403 &  1.548 \\
    BCDMS$_{120}$		& 0.35 - 0.75   & 13.25 - 57    & 32 & 36     & 50.913 &  1.591 & 49.821 &  1.383 \\
    BCDMS$_{200}$		& 0.35 - 0.75   & 32.50 - 99    & 28 & 28     & 63.344 &  2.262 & 37.020 &  1.322 \\
    BCDMS$_{280}$ 	& 0.35 - 0.75   & 43.00 - 115.5 & 26 & 26     & 29.949 &  1.151 & 23.490 &  0.903  \\
    NMC					    & 0.35 - 0.50   & 4.500 - 65.0  & 14 & 20     & 10.123 &  0.723 & 13.007 &  0.650 \\
    SLAC				    & 0.30 - 0.90   & 4.000 - 31.31 & 57 & 316    & 43.981 &  0.771 & 219.19 &  0.693 \\
    H1					    & 0.40 - 0.65   & 200.0 - 20000 & 26 & 26     & 37.372 &  1.437 & 36.770 &  1.414 \\
    ZEUS 				    & 0.40 - 0.65   & 650.0 - 30000 & 15 & 15     & 2.8309 &  0.188 & 2.7382 &  0.182 \\
    \midrule
    \bf F$\bf_2^{d}$\\
    BCDMS$_{120}$		& 0.35 - 0.75   & 13.25 - 57    & 32 & 36     & 55.084 &  1.721 & 70.657 &  1.962 \\
    BCDMS$_{200}$		& 0.35 - 0.75   & 32.50 - 99    & 28 & 28     & 41.659 &  1.487 & 26.418 &  0.943  \\
    BCDMS$_{280}$		& 0.35 - 0.75   & 43.00 - 115.5 & 26 & 26     & 28.072 &  1.079 & 15.393 &  0.592 \\
    NMC					    & 0.35 - 0.50   & 4.500 - 65.0  & 14 & 20     & 21.724 &  1.551 & 22.128 &  1.106 \\
    SLAC				    & 0.30 - 0.89   & 4.010 - 30.2  & 59 & 327    & 71.655 &  1.214 & 227.67 &  0.696 \\
    \midrule
    \bf F$\bf_2^{\bf NS}$\\
    BCDMS$_{120}$		& 0.0700 - 0.275 & 8.75 - 43    & 30 & 30     & 36.210 &  1.207 & 33.948 &  1.131 \\
    BCDMS$_{200}$		& 0.0700 - 0.275 & 17.0 - 75    & 28 & 28     & 20.120 &  0.718 & 20.358 &  0.727 \\
    BCDMS$_{280}$		& 0.1000 - 0.275 & 32.5 - 115.5 & 26 & 26     & 12.946 &  0.497 & 12.579 &  0.483 \\
    NMC					    & 0.0125 - 0.275 & 4.50 - 65    & 53 & 53     & 12.263 &  0.231 & 11.392 &  0.214 \\
    SLAC				    & 0.1530 - 0.293 & 4.18 - 8.22  & 28 & 28     & 24.138 &  0.862 & 21.323 &  0.761 \\
    \midrule
    \bf Total Data		& &  & 814 & 1395 &  &  \\
    \addlinespace
    \bf Total $\mathbf{\chi^2}$		& & & & & \multicolumn{2}{l}{814.188} & \multicolumn{2}{l}{1054.358} \\
    \addlinespace
    \bf Total $\mathbf{\chi^2/d.o.f.}$	& &	&  &  &  \multicolumn{2}{l}{814.188/806 = 1.010} & \multicolumn{2}{l}{1054.358/1377 = 0.765} \\
    \bottomrule
    \bottomrule
   \end{tabular}
   }
  \caption{\small
  Various combinations of the different subsets of $xF_3$ and non-singlet $F_2$ data, along with their corresponding $x$ and $Q^2$ ranges.
  The fourth and fifth columns present the number of individual data points for each data set considering $W^2\geq$12.5 GeV$^2$ and $W^2\geq$ 4 GeV$^2$ cuts on the data, respectively. Furthermore, the reduction in the number of CCFR data points due to additional cuts applied  to this data set (i.e., $x>0.4$) accounts for the discrepancy between CCFR and NuTeV in this region. The sixth and seventh columns ($\mathtt{SK24[QCD]}$) and the eighth and ninth ($\mathtt{SK24[QCD+TMC+HT]}$) contain the $\chi^2$ and $\chi^2/$\#Data values  for each fit applied on corresponding cuts. The total $\chi^2$ and $\chi^2/d.o.f.$ are also shown.
  }
 \label{tab:deltachi2fitted}
\end{table*}
\endgroup

\begingroup
\squeezetable

\begin{table*}[htp]
	\footnotesize
	\begin{tabular}{l@{\hskip 0.75in}l@{\hskip 0.75in}ll}
		\toprule
		\toprule
		Parameters & $\mathtt{SK24[QCD]}$ & $\mathtt{SK24[QCD+TMC+HT]}$ \\
		\midrule
		\midrule
		$N_u$		& $0.8426$               & $1.8777$\\
		$a_u$		& $0.4484   ~\pm~0.0265$ & $0.5626  ~\pm~ 0.0324$\\
		$b_u$		& $3.9178   ~\pm~0.0183$ & $4.2169  ~\pm~ 0.0185$\\
		$c_u$		& $9.3388   ~\pm~1.9361$ & $6.8458  ~\pm~ 1.1963$\\
		$d_u$		& $2.6550$               & $0.0$\\
		\midrule
		$N_d$		    & $ 1.5705$              & $4.9200$\\ 
		$a_d$		    & $ 0.5499  ~\pm~0.0149$ & $0.7715 ~ \pm~ 0.0201$\\
		$b_d$		    & $ 6.4429  ~\pm~0.1535$ & $7.0218 ~ \pm~ 0.1803$\\
		$c_d$		    & $ 8.9695$              & $6.7446 $\\ 
		$d_d$		    & $-2.072$               & $-2.957$\\
 		$e_d$	      & $0.0$                  & $0.4955  ~\pm~ 0.0731$\\
 		$f_d$	      & $0.0$                  & $6.3712 $\\
		\midrule
		$\alpha_s(M_Z^2)$ & $0.1154 ~ \pm~0.0009$ & $0.1149 ~ \pm~ 0.0014$ \\
		\midrule
		$h_{11}$		& $\qquad ~~~$ ... & $-11.71     \hspace{0.7pt} \pm~ 1.5808$\\
		$h_{12}$		& $\qquad ~~~$ ... & $~~~\!3.296 \hspace{0.5pt} \pm~ 0.2311$\\
		$h_{13}$		& $\qquad ~~~$ ... & $-1.868     \hspace{0.9pt} \pm~ 0.1114$\\
		$h_{21}$		& $\qquad ~~~$ ... & $-2.005     \hspace{0.9pt} \pm~ 0.7045$\\
		$h_{22}$		& $\qquad ~~~$ ... & $~~~\!2.584 \hspace{0.5pt}  \pm~ 0.2172$\\
		$h_{23}$		& $\qquad ~~~$ ... & $-7.203     \hspace{0.9pt} \pm~ 2.2586$\\
		$h_{31}$		& $\qquad ~~~$ ... & $-12.69     \hspace{0.9pt}   \pm~ 3.4884$\\
		$h_{32}$		& $\qquad ~~~$ ... & $~~~\!1.641 \hspace{0.5pt} \pm~ 0.2168$\\
		$h_{33}$		& $\qquad ~~~$ ... & $-1.952     \hspace{0.9pt} \pm~ 0.1035$\\		
    \midrule
    $C_{N}$		  & $0.4468    \hspace{3pt} \pm~0.0949$ & $0.4938    \hspace{3pt} \pm~0.1096$\\
    $x_{0}$		  & $0.8142    \hspace{3pt} \pm~0.0696$ & $0.5491    \hspace{3pt} \pm~0.0451$\\
    $x_{1}$		  & $0.0986$                            & $0.0856$                           \\
		\bottomrule
		\bottomrule
	\end{tabular}
	\caption{The parameter values of the $u$- and $d$-valence quark densities in Eqs.~(\ref{eq:parm1}, \ref{eq:parm2}) at the input scale of $Q^2 = 4.0 $ GeV$^2$, obtained from the best fit considering QCD and nuclear corrections ($\mathtt{SK24[QCD]}$), as well as TMC and HT corrections ($\mathtt{SK24[QCD+TMC+HT]}$) at NNLO. The parameter values without error have been fixed after the first minimization, due to the fact that the data do not constrain some parameters well enough. 
	}
	\label{tab:parm}
\end{table*}
\endgroup

\begingroup
\squeezetable

\begin{table*}[htp]
	\footnotesize
	\resizebox{\textwidth}{!}{
	\begin{tabular}{|r|r|r|r|r|r|r|r|r|}
		\hline 
    \hline 
		{\bf NNLO} &\multicolumn{1}{|c|}{$a_u$}      & \multicolumn{1}{|c|}{$b_u$}      & \multicolumn{1}{|c|}{$c_u$}      & \multicolumn{1}{|c|}{$a_d$}      & \multicolumn{1}{|c|}{$b_d$}      & \multicolumn{1}{|c|}{$\alpha_s$} & \multicolumn{1}{|c|}{$C_{N}$}    & \multicolumn{1}{|c|}{$x_{0}$}   \\
		\hline 
    \hline 
     \multicolumn{1}{|c|}{$a_u$}      &  \textbf{0.704E-03} &                &                &                &                &                &                &              \\ \hline
     \multicolumn{1}{|c|}{$b_u$}      & $-0.282$E-$03$ & \textbf{0.338E-03} &                &                &                &                &                &              \\ \hline
     \multicolumn{1}{|c|}{$c_u$}      & $-0.508$E-$01$ & $ 0.233$E-$01$ & \textbf{0.375E+01} &                &                &                &                &              \\ \hline
     \multicolumn{1}{|c|}{$a_d$}      & $-0.160$E-$03$ & $ 0.384$E-$05$ & $ 0.123$E-$01$ & \textbf{0.223E-03} &                &                &                &              \\ \hline
     \multicolumn{1}{|c|}{$b_d$}      & $-0.202$E-$02$ & $ 0.616$E-$03$ & $ 0.157$E+$00$ & $ 0.194$E-$02$ & \textbf{0.236E-01} &                &                &              \\ \hline
     \multicolumn{1}{|c|}{$\alpha_s$} & $ 0.972$E-$05$ & $-0.603$E-$05$ & $-0.584$E-$03$ & $ 0.357$E-$05$ & $-0.241$E-$05$ & \textbf{0.844E-06} &                &              \\ \hline
     \multicolumn{1}{|c|}{$C_{N}$}    & $ 0.324$E-$03$ & $-0.552$E-$03$ & $-0.323$E-$01$ & $-0.365$E-$03$ & $-0.245$E-$02$ & $-0.108$E-$04$ & \textbf{0.901E-02} &              \\ \hline
     \multicolumn{1}{|c|}{$x_{0}$}    & $-0.448$E-$03$ & $ 0.496$E-$03$ & $ 0.389$E-$01$ & $ 0.379$E-$03$ & $ 0.431$E-$02$ & $ 0.664$E-$06$ & $-0.617$E-$02$ & \textbf{0.485E-02} \\ \hline
		\hline 
	\end{tabular}}
	\caption{NNLO covariance matrix of $\mathtt{SK24[QCD]}$ fit at Q$_0^2= 4$ GeV$^2$. }
	\label{tab:QCDcov}
\end{table*}
\endgroup

\begin{table*}[htb]
  \resizebox{1.0\textwidth}{!}{
\begin{tabular}{c|c|c|c|c|c|c|c|c|c}
\hline\hline  $<x^{n-1}>_{q_{v}}$ & $\mathtt{SK24[QCD]}$& $\mathtt{SK24[QCD+TMC+HT]}$ & $\mathtt{GKA19}$ & $\mathtt{KT08}$ & $\mathtt{KT07}$ & $\mathtt{BBG06}$ &$\mathtt{MSHT20}$ & $\mathtt{NNPDF4.0}$ &$\mathtt{CT18}$ \\
  &&&(Jacobi poly.)  & (Bernstein poly.) &&&&\\
\hline   
   $<x^1>_{u_{v}}$    & $0.3009\pm 0.0063$  & $0.3011 \pm 0.0136$  & $0.3112$ & $0.3056$ & $0.2934$ & $0.2986$  & $0.2855$ & $0.2845$ & $0.2902$\\
   $<x^2>_{u_{v}}$    & $0.0887\pm 0.0018$  & $0.0876 \pm 0.0041$  & $0.0890$ & $0.0871$ & $0.0825$ & $0.0871$  & $0.0834$ & $0.0820$ & $0.0845$\\
   $<x^3>_{u_{v}}$    & $0.0344\pm 0.0007$  & $0.0335 \pm 0.0017$  & $0.0340$ & $0.0330$ & $0.0311$ & $0.0333$  & $0.0324$ & $0.0313$ & $0.0327$\\
\hline   
  $<x^1>_{d_{v}}$     & $0.1024\pm 0.0068$  & $0.1131 \pm 0.0135$ & $0.1019$ & $0.1235$ & $0.1143$ & $0.1239$  & $0.1147$ & $0.1099$ & $0.1200$\\
  $<x^2>_{d_{v}}$     & $0.0231\pm 0.0020$  & $0.0274 \pm 0.0045$ & $0.0207$ & $0.0298$ & $0.0262$ & $0.0315$  & $0.0289$ & $0.0267$ & $0.0300$\\
  $<x^3>_{d_{v}}$     & $0.0071\pm 0.0008$  & $0.0094 \pm 0.0020$ & $0.0058$ & $0.0098$ & $0.0083$ & $0.0105$  & $0.0097$ & $0.0086$ & $0.0103$\\
\hline
  $<x^1>_{u_v - d_v}$ & $0.1985\pm 0.0104$  & $0.1880 \pm 0.0186$ & $0.2093$ & $0.1821$ & $0.1791$ & $0.1747$  & $0.1708$ & $0.1746$ & $0.1701$\\
  $<x^2>_{u_v - d_v}$ & $0.0655\pm 0.0030$  & $0.0602 \pm 0.0060$ & $0.0683$ & $0.0573$ & $0.0563$ & $0.0556$  & $0.0545$ & $0.0553$ & $0.0545$\\
  $<x^3>_{u_v - d_v}$ & $0.0273\pm 0.0012$  & $0.0240 \pm 0.0026$ & $0.0282$ & $0.0232$ & $0.0228$ & $0.0228$  & $0.0227$ & $0.0227$ & $0.0224$\\
\hline\hline
\end{tabular}}
\caption{Low order moments ($n=2,3,4$)  of $<x^{n-1}>_{u_v,d_v,u_v-d_v}$, at  $Q^2$= 4 GeV$^2$ for our QCD non-singlet fits ($\mathtt{SK24[QCD]}$)
with comparison of the NNLO analysis, GKA19~\cite{Tooran:2019cfz}, KT08~\cite{Khorramian:2008yh}, KT07~\cite{Khorramian:2006wg}, BBG06~\cite{Blumlein:2006be},
MSHT20~\cite{Bailey:2020ooq}, NNPDF4.0 \cite{NNPDF:2021njg} and
CT18~\cite{Hou:2019qau}.}
\label{tab:LowMom}
\end{table*}

\clearpage


\begin{figure*}[h!]

\includegraphics[width=0.6\textwidth]{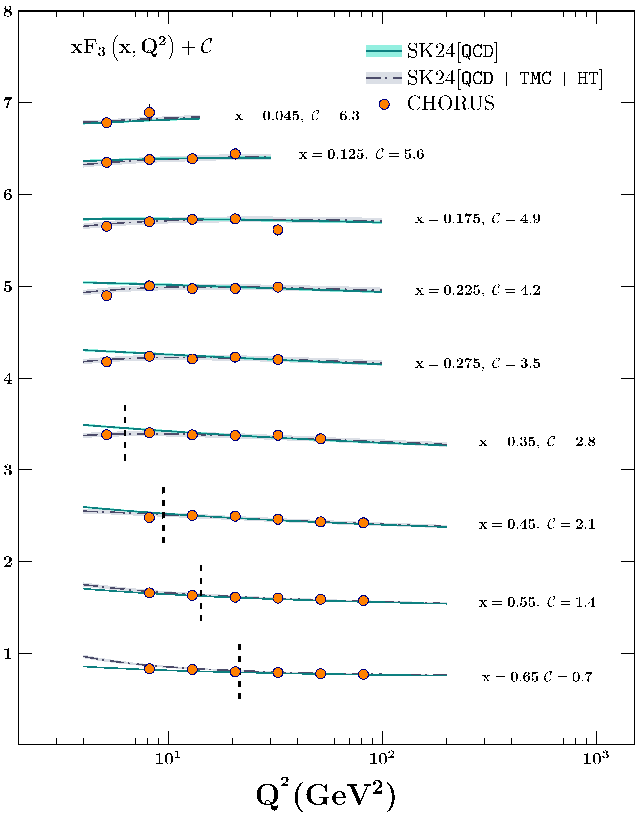}
  
\caption{The comparison of structure function $xF_3$ (Lead target) obtained from the $\mathtt{SK24[QCD]}$ and $\mathtt{SK24[QCD+TMC+HT]}$ fits  as a function of $Q^2$ in the various $x$, at the NNLO approximation. The vertical dashed line indicates the regions with W$^2$ $\geq$ 12.5$~$GeV$^2$.}
\label{fig:All-In-One-xF3Pb}
\end{figure*}
\begin{figure*}[h!]

\includegraphics[width=0.6\textwidth]{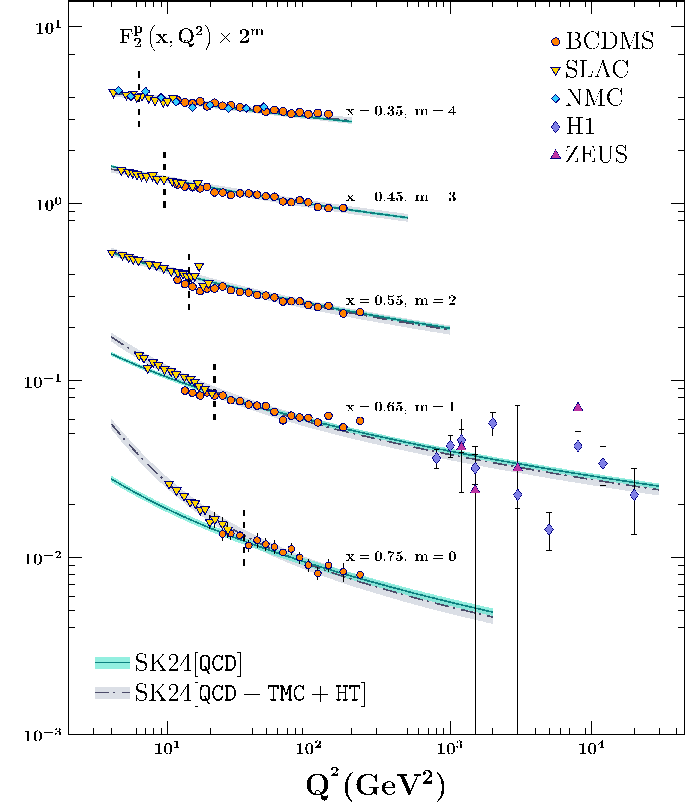}
  
\caption{The comparison of structure function $F_2^p$ obtained from the $\mathtt{SK24[QCD]}$ and $\mathtt{SK24[QCD+TMC+HT]}$ fits  as a function of $Q^2$ in the various $x$, at the NNLO approximation. The vertical dashed line indicates the regions with W$^2$ $\geq$ 12.5$~$GeV$^2$.
 }
\label{fig:All-In-One-F2P}
\end{figure*}
\begin{figure*}[h!]

\includegraphics[width=0.6\textwidth]{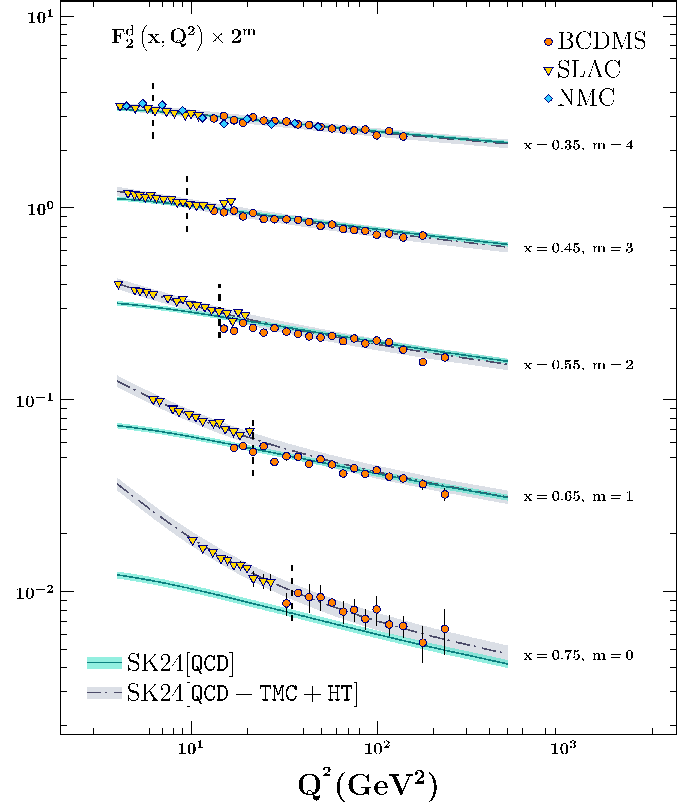}
  
\caption{The comparison of structure function $F_2^d$ obtained from the $\mathtt{SK24[QCD]}$ and $\mathtt{SK24[QCD+TMC+HT]}$ fits  as a function of $Q^2$ in the various $x$, at the NNLO approximation. The vertical dashed line indicates the regions with W$^2$ $\geq$ 12.5$~$GeV$^2$.
 }
\label{fig:All-In-One-F2D}
\end{figure*}
\begin{figure*}[h!]

\includegraphics[width=0.6\textwidth]{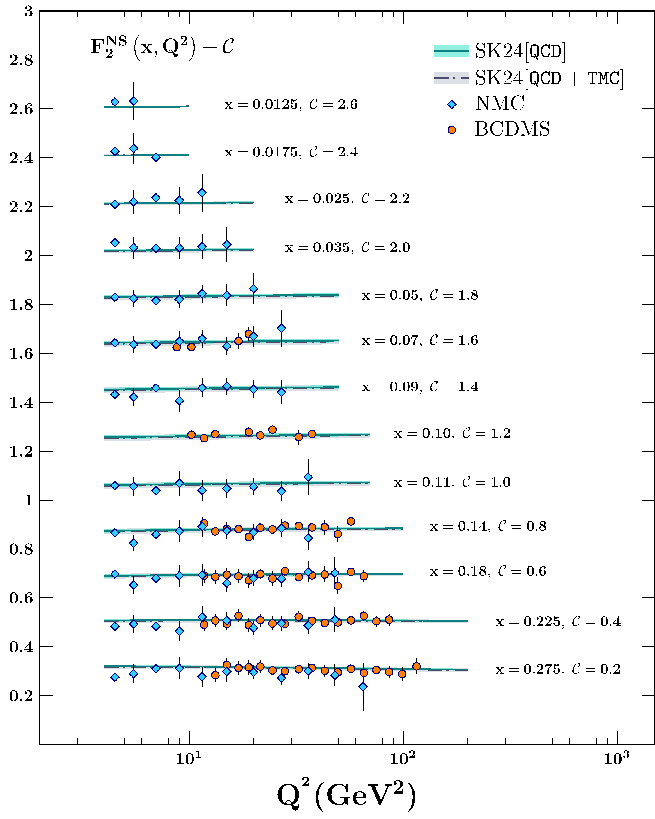}
  
\caption{The comparison of structure function $F_2^{NS}$ obtained from the $\mathtt{SK24[QCD]}$ and $\mathtt{SK24[QCD+TMC+HT]}$ fits  as a function of $Q^2$ in the various $x$, at the NNLO approximation.
 }
\label{fig:All-In-One-F2NS}
\end{figure*}
\begin{figure*}[h!]

   \includegraphics[width=0.3\textwidth]{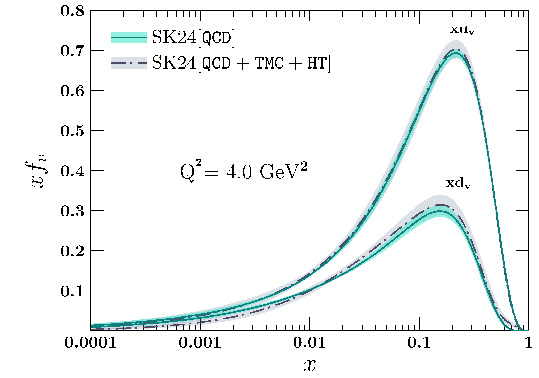}
   \includegraphics[width=0.287\textwidth]{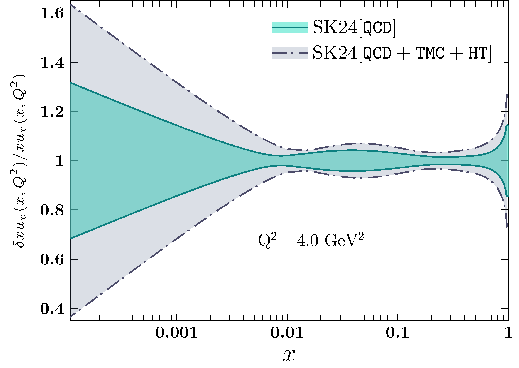}
   \includegraphics[width=0.290\textwidth]{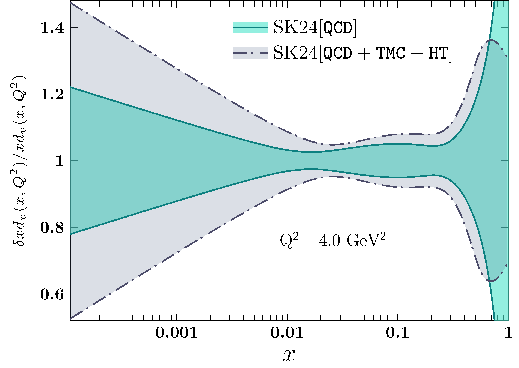}\\
   \includegraphics[width=0.3\textwidth]{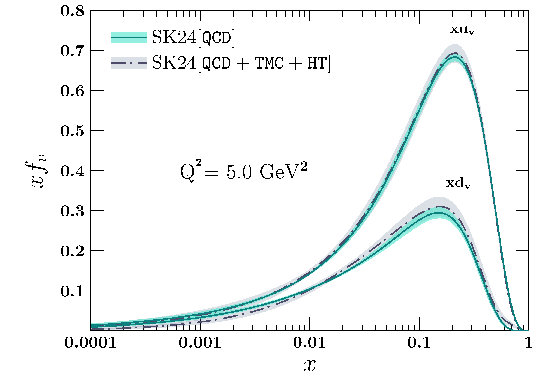}
   \includegraphics[width=0.287\textwidth]{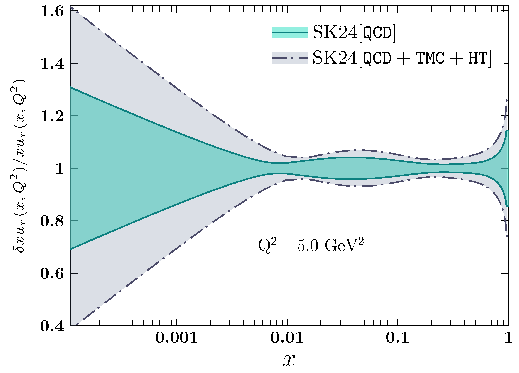}
   \includegraphics[width=0.290\textwidth]{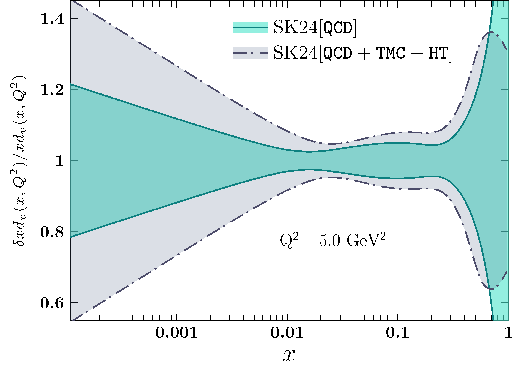}\\
   \includegraphics[width=0.3\textwidth]{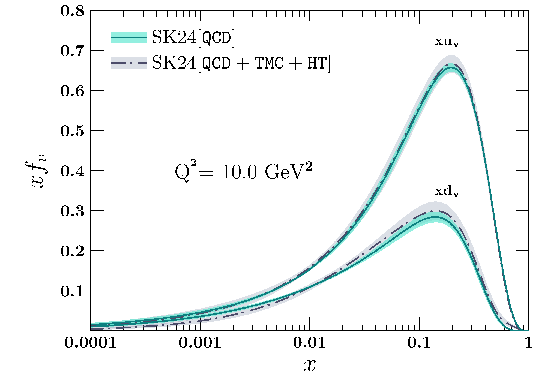}
   \includegraphics[width=0.287\textwidth]{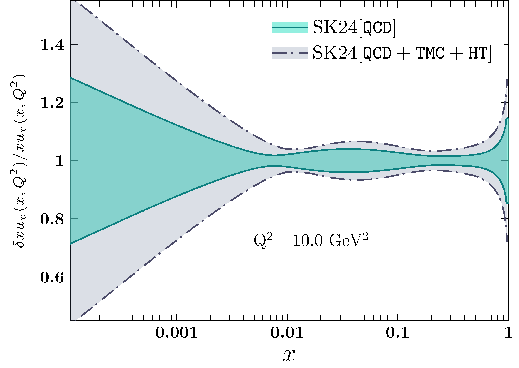}
   \includegraphics[width=0.290\textwidth]{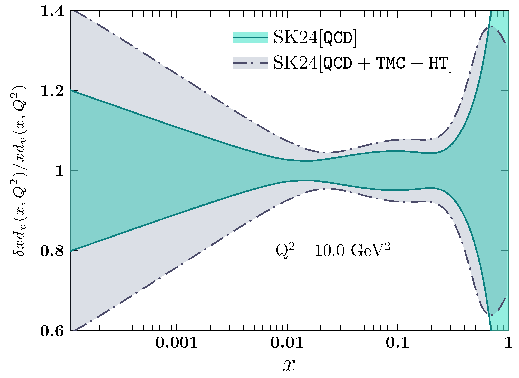}\\
   \includegraphics[width=0.3\textwidth]{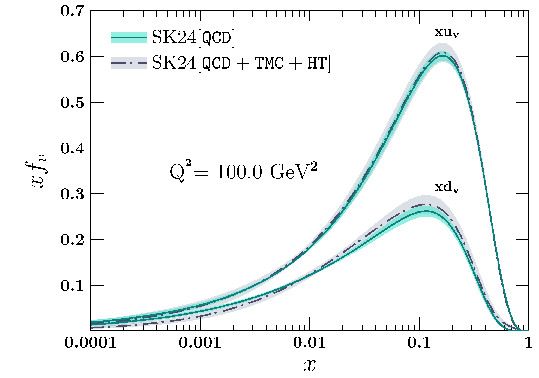}
   \includegraphics[width=0.287\textwidth]{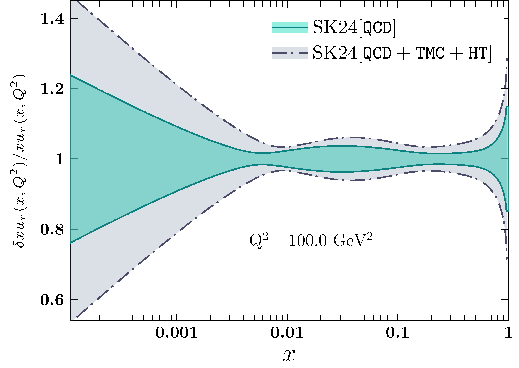}
   \includegraphics[width=0.290\textwidth]{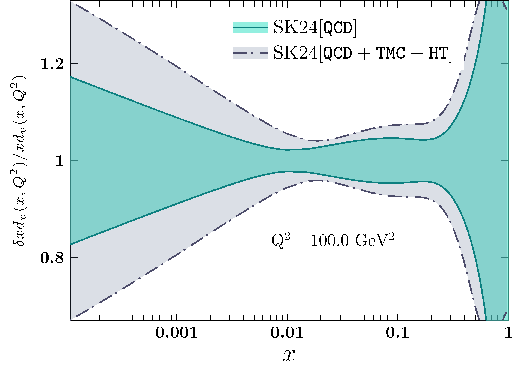}\\
   \includegraphics[width=0.3\textwidth]{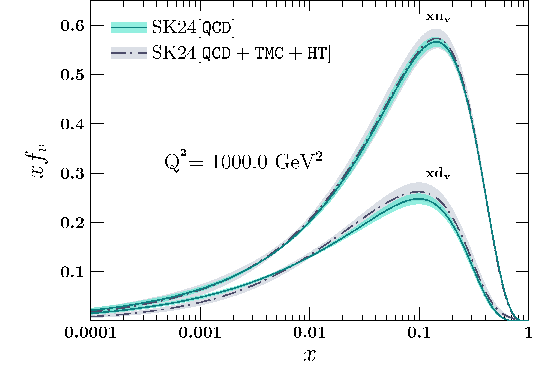}
   \includegraphics[width=0.287\textwidth]{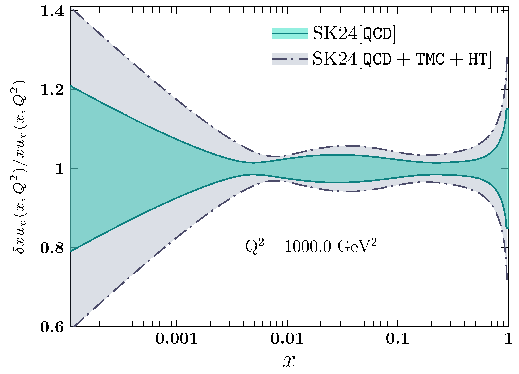}
   \includegraphics[width=0.290\textwidth]{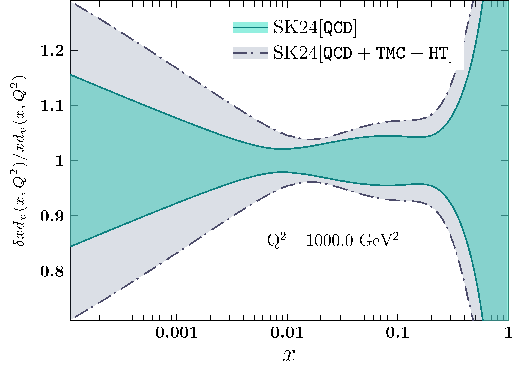}\\
   \includegraphics[width=0.3\textwidth]{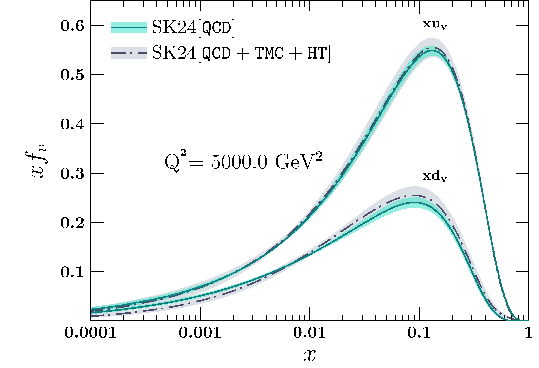}
   \includegraphics[width=0.287\textwidth]{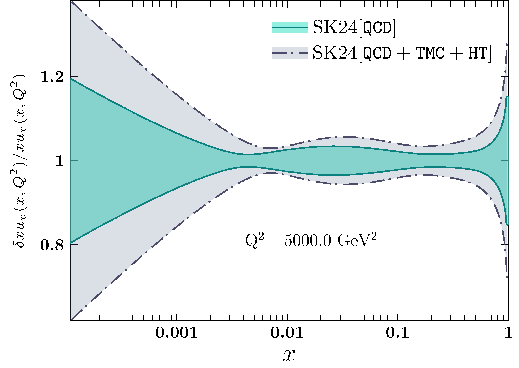}
   \includegraphics[width=0.290\textwidth]{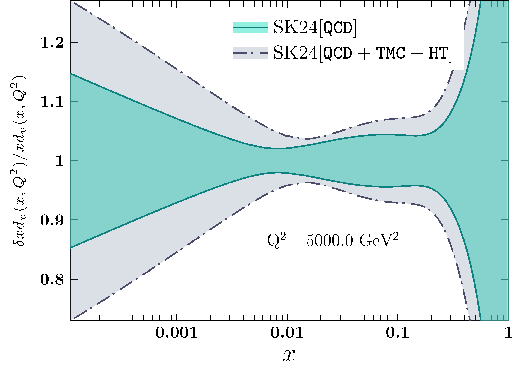}
    
  \caption{The NNLO comparison of $xu_v$ and $xd_v$ obtained from the $\mathtt{SK24[QCD]}$ and $\mathtt{SK24[QCD+TMC+HT]}$ at $Q^2$=4,~5,~10,~100,~1000,~5000 GeV$^2$ as a function of $x$. In the left panels, our results for $\mathtt{SK24[QCD]}$ and $\mathtt{SK24[QCD+TMC+HT]}$ are presented, while the middle and right panels display the relative uncertainties $\delta xu_v(x,Q^2)/xu_v(x,Q^2)$ and  $\delta xd_v(x,Q^2)/xd_v(x,Q^2)$, respectively.}
  \label{fig:QCD-HT PDFs}
  \end{figure*}
\begin{figure*}[h!]

  \includegraphics[width=0.245\textwidth]{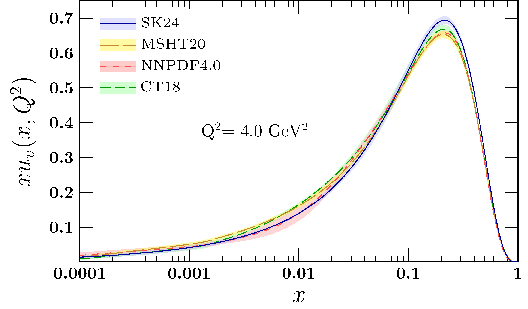}
  \includegraphics[width=0.245\textwidth]{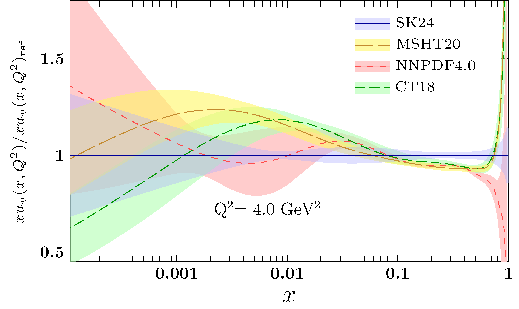}
  \includegraphics[width=0.245\textwidth]{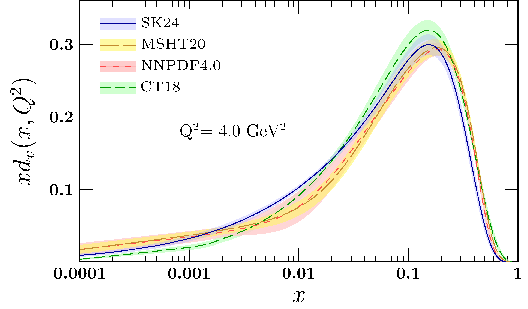}
  \includegraphics[width=0.245\textwidth]{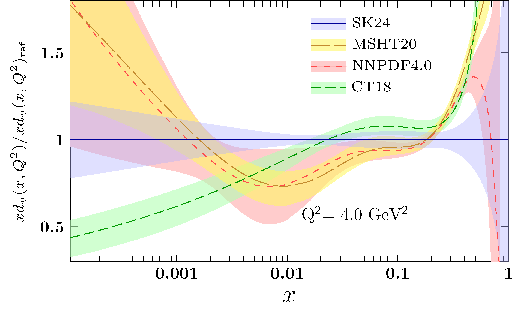}\\

  \includegraphics[width=0.245\textwidth]{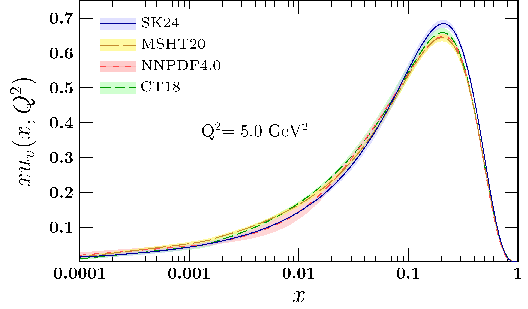}
  \includegraphics[width=0.245\textwidth]{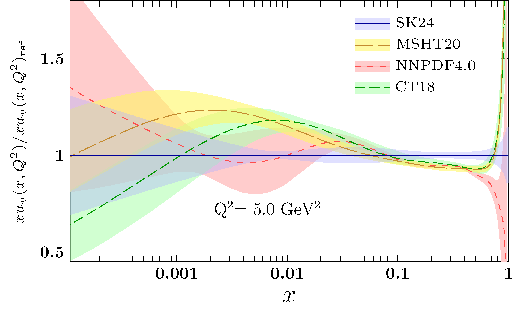}
  \includegraphics[width=0.245\textwidth]{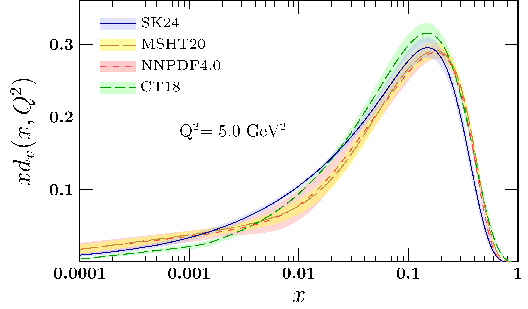}
  \includegraphics[width=0.245\textwidth]{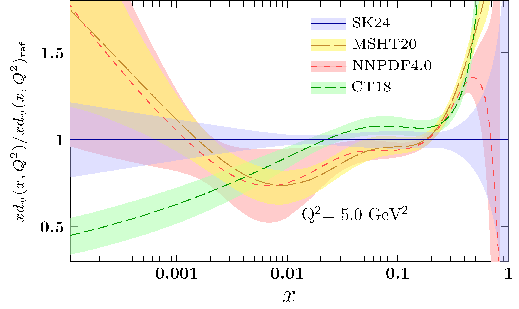}\\

  \includegraphics[width=0.245\textwidth]{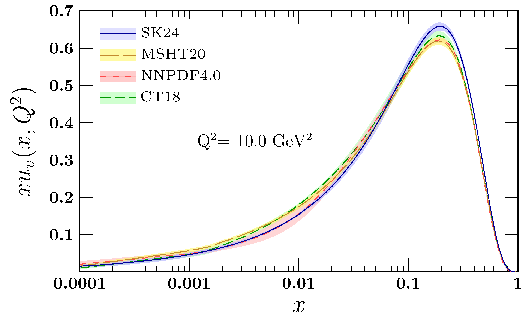}
  \includegraphics[width=0.245\textwidth]{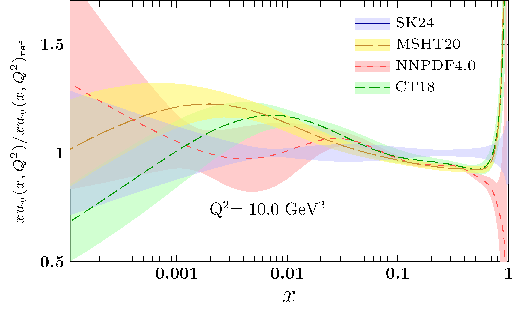}
  \includegraphics[width=0.245\textwidth]{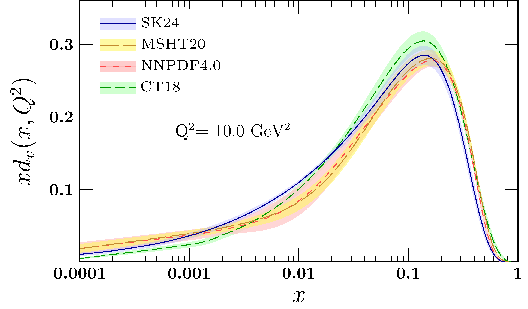}
  \includegraphics[width=0.245\textwidth]{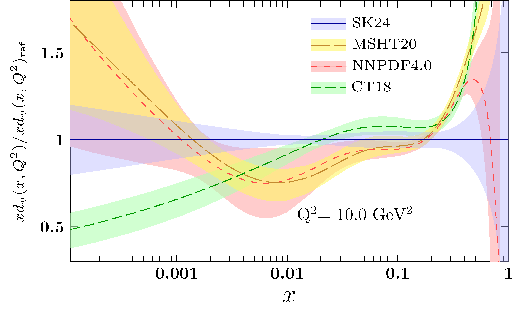}\\

  \includegraphics[width=0.245\textwidth]{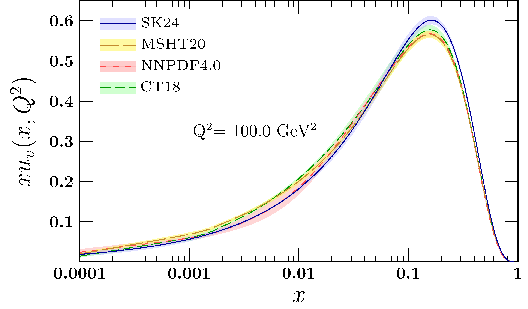}
  \includegraphics[width=0.245\textwidth]{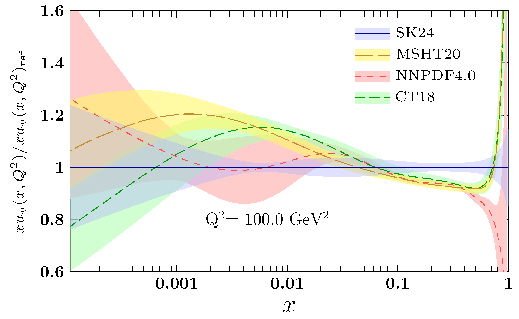}
  \includegraphics[width=0.245\textwidth]{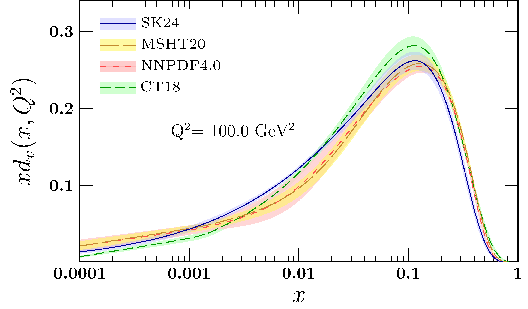}
  \includegraphics[width=0.245\textwidth]{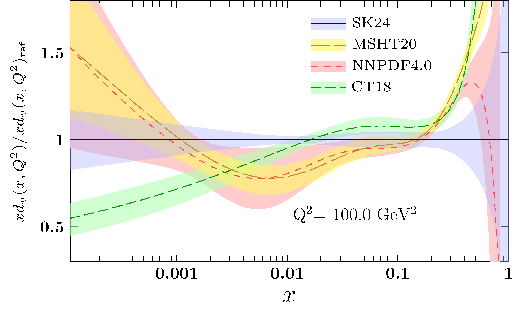}\\

  \includegraphics[width=0.245\textwidth]{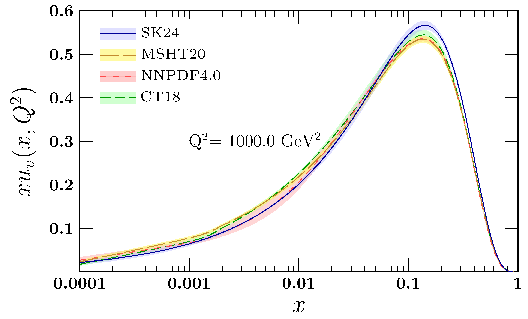}
  \includegraphics[width=0.245\textwidth]{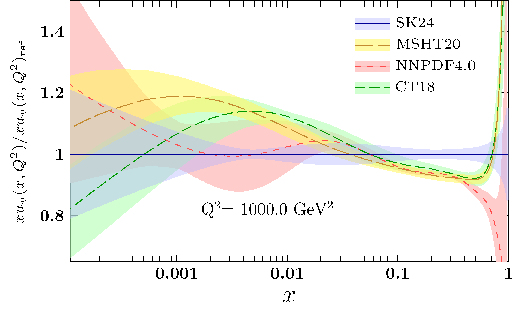}
  \includegraphics[width=0.245\textwidth]{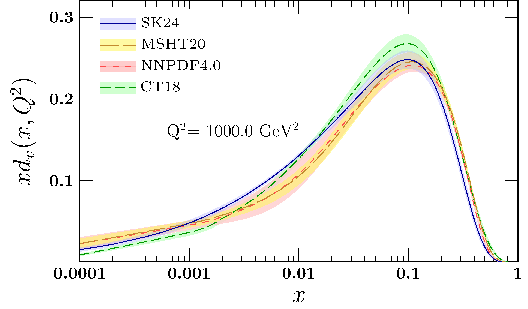}
  \includegraphics[width=0.245\textwidth]{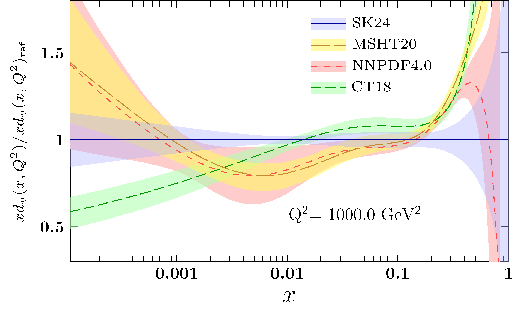}\\

  \includegraphics[width=0.245\textwidth]{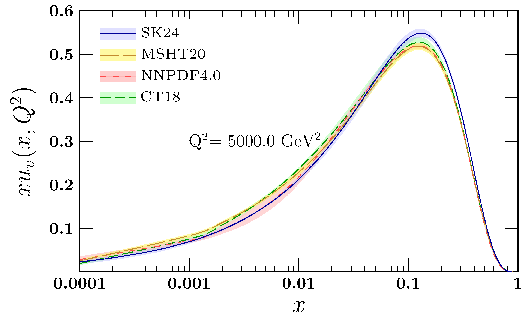}
  \includegraphics[width=0.245\textwidth]{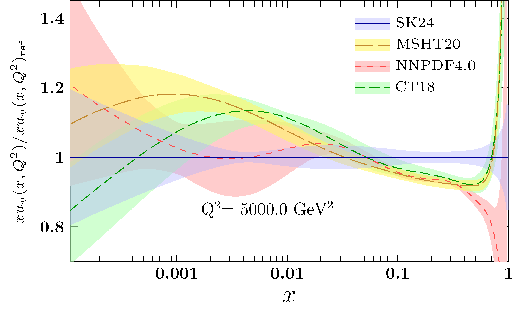}
  \includegraphics[width=0.245\textwidth]{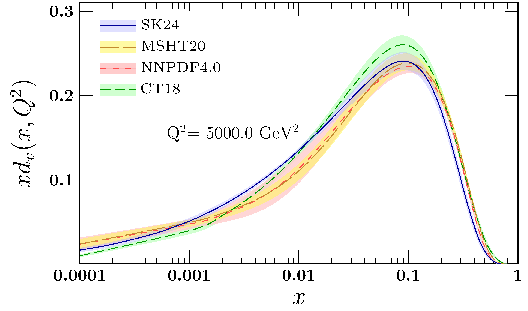}
  \includegraphics[width=0.245\textwidth]{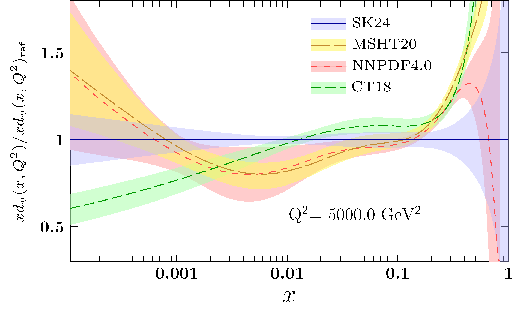}

\caption{The NNLO comparison of $xu_v(x,Q^2)$ and $xd_v(x,Q^2)$ obtained from the $\mathtt{SK24[QCD]}$ fit  compared with MSHT20 \cite{Bailey:2020ooq}, NNPDF4.0 \cite{NNPDF:2021njg} and CT18 \cite{Hou:2019qau} non-singlet/singlet predictions  at $Q^2$=4,~5,~10,~100,~1000,~5000 GeV$^2$ as a function of $x$. In the first and third columns our results for $\mathtt{SK24[QCD]}$ are presented, and the second and fourth columns show log plots for ratios of $xu_v(x,Q^2)/xu_v(x,Q^2)_{ref}$ and $xd_v(x,Q^2)/xd_v(x,Q^2)_{ref}$, with respect to $\mathtt{SK24[QCD]}$.}
\label{fig:PDFs-xuv-xdv}
\end{figure*}
\begin{figure*}[h!]

  \includegraphics[width=0.32\textwidth]{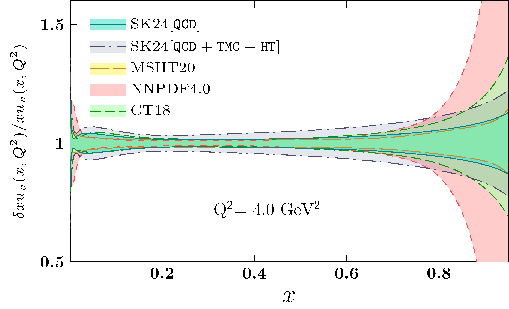}
  \includegraphics[width=0.32\textwidth]{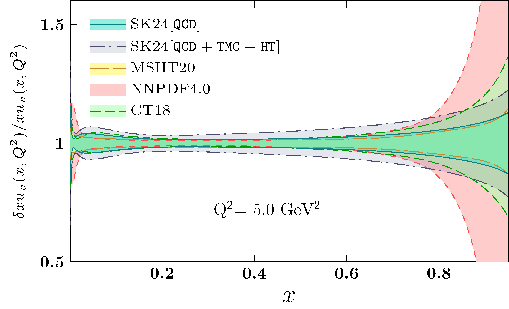}
  \includegraphics[width=0.32\textwidth]{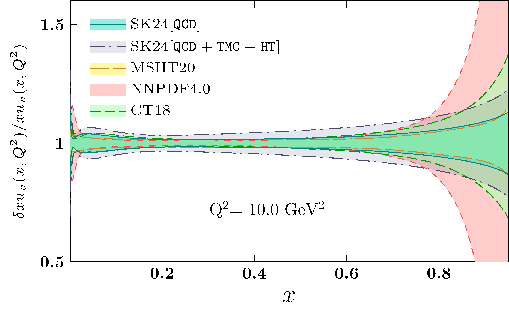}\\
  \includegraphics[width=0.32\textwidth]{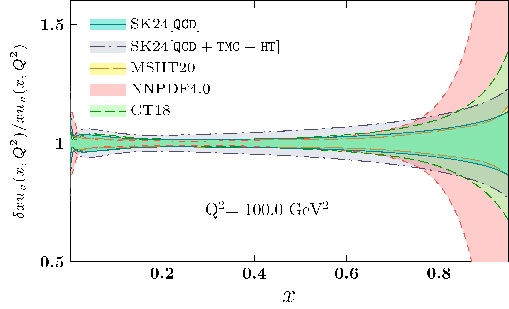}
  \includegraphics[width=0.32\textwidth]{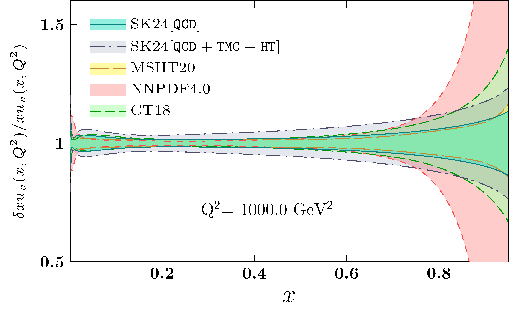}
  \includegraphics[width=0.32\textwidth]{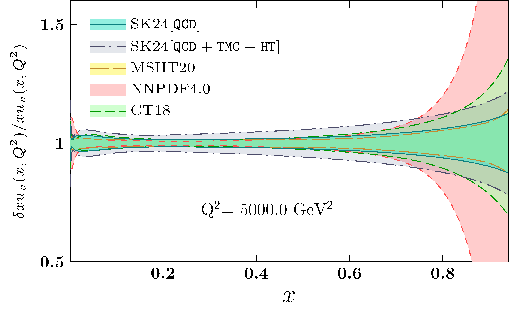}

  \caption{The NNLO comparison of relative uncertainties $\delta xu_v(x,Q^2)/xu_v(x,Q^2)$ obtained from the $\mathtt{SK24[QCD]}$ and $\mathtt{SK24[QCD+TMC+HT]}$ fits as a function of $x$ at $Q^2$=4,~5,~10,~100,~1000,~5000 GeV$^2$ compared with MSHT20 \cite{Bailey:2020ooq}, NNPDF4.0 \cite{NNPDF:2021njg} and CT18 \cite{Hou:2019qau} non-singlet/singlet predictions.}
\label{fig:xuv-err}
\end{figure*}
\begin{figure*}[h!]

  \includegraphics[width=0.32\textwidth]{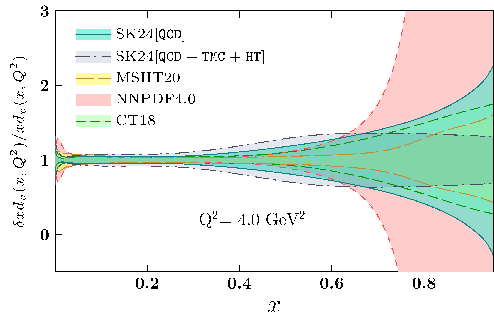}
  \includegraphics[width=0.32\textwidth]{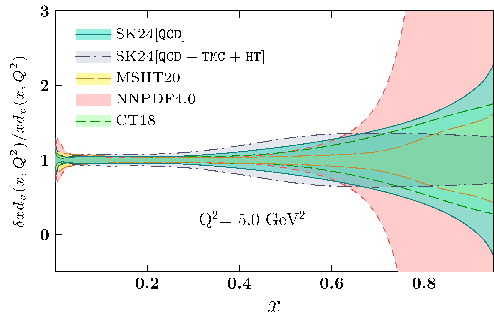}
  \includegraphics[width=0.32\textwidth]{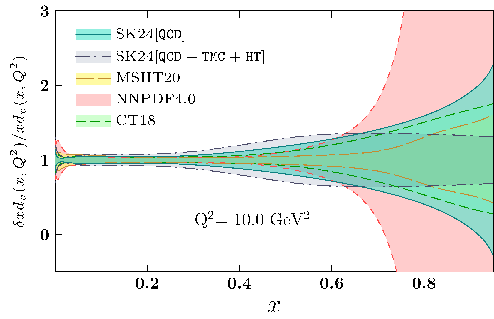}\\
  \includegraphics[width=0.32\textwidth]{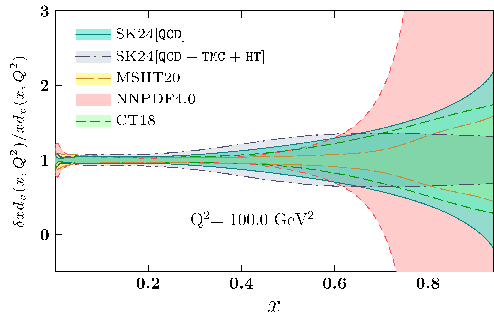}
  \includegraphics[width=0.32\textwidth]{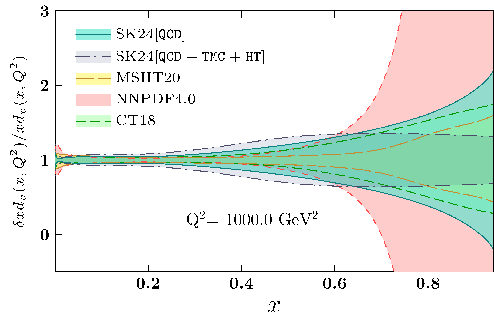}
  \includegraphics[width=0.32\textwidth]{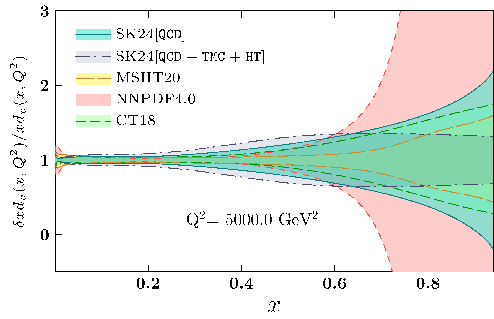}

  \caption{The NNLO comparison of relative uncertainties $\delta xd_v(x,Q^2)/xd_v(x,Q^2)$ obtained from the $\mathtt{SK24[QCD]}$ and $\mathtt{SK24[QCD+TMC+HT]}$ fits as a function of $x$ at $Q^2$=4,~5,~10,~100,~1000,~5000 GeV$^2$ compared with MSHT20 \cite{Bailey:2020ooq}, NNPDF4.0 \cite{NNPDF:2021njg} and CT18 \cite{Hou:2019qau} non-singlet/singlet predictions.}
\label{fig:xdv-err}
\end{figure*}
 
\begin{figure*}[h!]

  \includegraphics[width=0.95\textwidth]{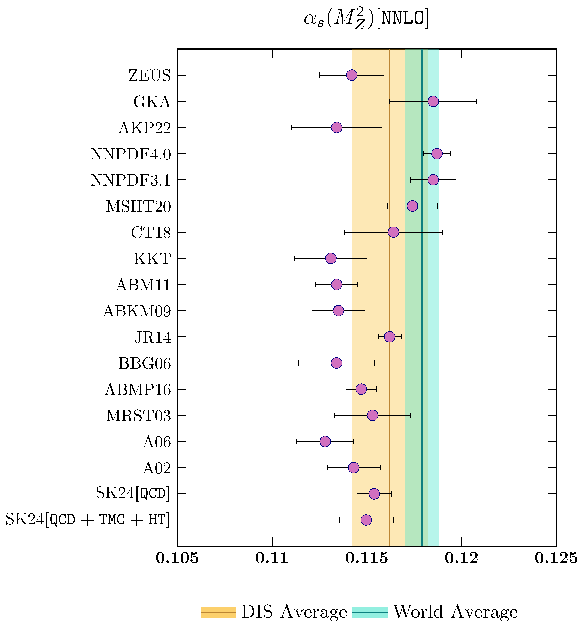}
   
  \caption{The obtained values and uncertainties of $\alpha_{s}(M_{Z}^{2})$ in present non-singlet $\mathtt{SK24[QCD]}$ and $\mathtt{SK24[QCD+TMC+HT]}$ analysis comparing with available results A02~\cite{Alekhin:2002fv}, A06~\cite{Alekhin:2006zm}, MRST03~\cite{Martin:2003tt}, ABMP16~\cite{Alekhin:2017kpj, Alekhin:2018pai}, BBG06~\cite{Blumlein:2006be}, JR14~\cite{Jimenez-Delgado:2014twa}, ABKM09~\cite{Alekhin:2009ni}, ABM11~\cite{Alekhin:2012ig}, KKT~\cite{Khorramian:2009xz}, CT18~\cite{Hou:2019qau}, MSHT20~\cite{Bailey:2020ooq}, NNPDF3.1~\cite{Ball:2018iqk}, NNPDF4.0~\cite{NNPDF:2021njg}, AKP22~\cite{Azizi:2022nqm}, GKA~\cite{Tooran:2019cfz} and ZEUS~\cite{ZEUS:2023zie} at the NNLO approximation. Furthermore, the world and DIS average values of the strong coupling constant $\alpha_{s}(M_{Z}^{2})$  ~\cite{ParticleDataGroup:2022pth} illustrated for well comparison.}
  \label{Fig:AlphaS}
 \end{figure*}
\begin{figure*}[h!]
  
 \includegraphics[width=0.75\textwidth]{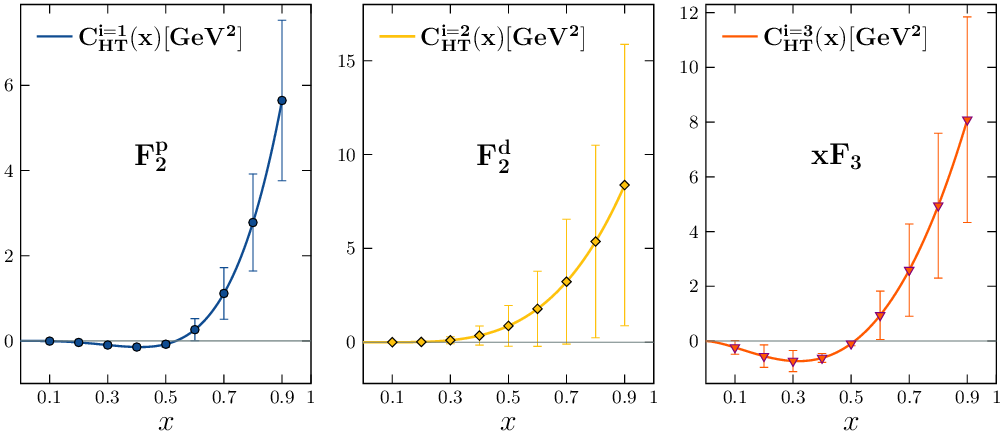}
  
\caption{Impact of higher twist function ${C_{HT}^{i} (x) [GeV^2]}$ from Eq. \eqref{eq:HTform} on large $x$ region for $i=1, 2,3$ corresponding to the HT effects for the $F_2^{p}$, $F_2^{d}$, and $xF_3$ structure functions, respectively (see text).}

\label{fig:HTs}
\end{figure*}
\begin{figure*}[h!]

 \includegraphics[width=0.40\textwidth]{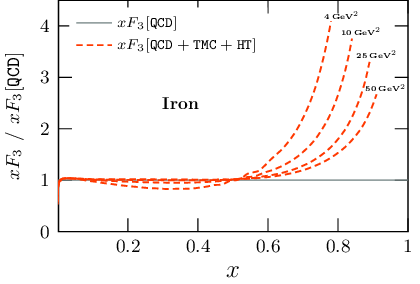}\\
 \includegraphics[width=0.40\textwidth]{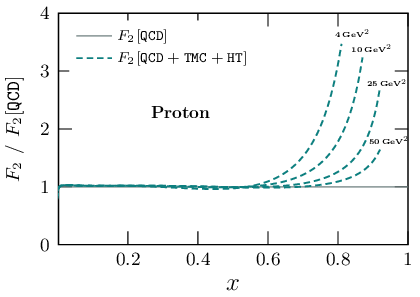}\\
 \includegraphics[width=0.40\textwidth]{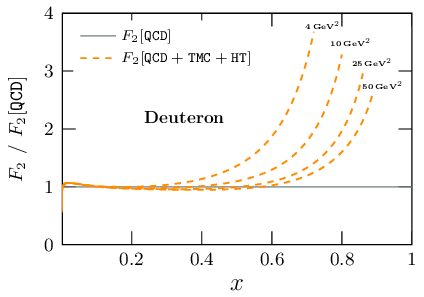}\\
  
\caption{Ratio of the $xF_3/xF_3[\mathtt{QCD}]$, $F_2^p/F_2^p[\mathtt{QCD}]$, and $F_2^d/F_2^d[\mathtt{QCD}]$ as a function of $x$ for Q$^2$ = 4, 10, 25 and 50 GeV$^2$.}
\label{fig:TMC-Ratio}
\end{figure*}

\end{document}